\documentclass[journal,twocolumn,twoside,10pt,comsoc]{IEEEtran}

\usepackage[T1]{fontenc} 
\usepackage{amsmath}
\usepackage[cmintegrals]{newtxmath}

\usepackage[numbers,sort,compress]{natbib}

\usepackage{tikz}
\usetikzlibrary{automata,chains,shapes,decorations}

\usepackage{graphicx,pgfplots}
\usetikzlibrary{plotmarks}
\usepgfplotslibrary{fillbetween}
\pgfplotsset{compat=newest}
\pgfplotstableset{col sep=comma}

\usepackage{todonotes,comment,changes}

\usepackage{mathtools,mathrsfs}
\usepackage{nicefrac}
\usepackage[autolanguage]{numprint}
\usepackage[ruled, linesnumbered, noend]{algorithm2e}
\SetKw{Break}{break}
\SetKw{And}{and}
\SetKw{Or}{or}

\usepackage[caption=false, font=footnotesize]{subfig}

\usepackage[colorlinks=true, citecolor=blue]{hyperref}

\usepackage{booktabs}
\usepackage[group-digits=integer, group-separator={,}, group-minimum-digits = 4, detect-family=true, exponent-product=\cdot]{siunitx}

\newcommand{\midpoint}{\operatorname{mid}}

\newcommand{\acces}{\rightsquigarrow}
\newcommand{\upacces}{\rightharpoonup}

\begin{document}

\title{Imprecise Markov Models \\ for Scalable and Robust Performance Evaluation \\ of Flexi-Grid Spectrum Allocation Policies}

\author{
	Alexander~Erreygers, Cristina~Rottondi, Giacomo~Verticale and~Jasper~De~Bock
	\thanks{A shortened version of this work---not including Appendices~\ref{app:PreciseErgodicity} and \ref{app:LTROs}---has been submitted to the IEEE for possible publication.
	Copyright may be transferred without notice, after which this version may no longer be accessible.}%
	\thanks{A. Erreygers is a member of ELIS, Ghent University, 9052 Ghent, Belgium (email: alexander.erreygers@ugent.be).}%
	\thanks{C. Rottondi is with the Dalle Molle Institute for Artificial Intelligence, University of Lugano and University of Applied Science and Arts of Southern Switzerland, 6928 Manno, Switzerland (email: cristina.rottondi@supsi.ch).}%
	\thanks{G. Verticale is with the Dipartimento di Elettronica, Informazione e Bioingegneria, Politecnico di Milano, 20133 Milan, Italy (email: giacomo.verticale@polimi.it).}%
	\thanks{J. De Bock is a member of ELIS, Ghent University, 9052 Ghent, Belgium (email: jasper.debock@ugent.be).}%
}%


\maketitle

\begin{abstract}
	The possibility of flexibly assigning spectrum resources with channels of different sizes greatly improves the spectral efficiency of optical networks, but can also lead to unwanted spectrum fragmentation.
	We study this problem in a scenario where traffic demands are categorised in two types (low or high bit-rate) by assessing the performance of three allocation policies.
	Our first contribution consists of exact Markov chain models for these allocation policies, which allow us to numerically compute the relevant performance measures.
	However, these exact models do not scale to large systems, in the sense that the computations required to determine the blocking probabilities---which measure the performance of the allocation policies---become intractable.
	In order to address this, we first extend an approximate reduced-state Markov chain model that is available in the literature to the three considered allocation policies.
	These reduced-state Markov chain models allow us to tractably compute approximations of the blocking probabilities, but the accuracy of these approximations cannot be easily verified.
	Our main contribution then is the introduction of reduced-state imprecise Markov chain models that allow us to derive guaranteed lower and upper bounds on blocking probabilities, for the three allocation policies separately or for all possible allocation policies simultaneously.
\end{abstract}


\IEEEpeerreviewmaketitle

	\section{Introduction}
Flexi-grid optical networks \cite{gerstel2012elastic} have been envisioned as a novel paradigm to cope with the ever-growing Internet traffic: spectral resources are divided in small frequency slices (e.g., \SI{12.5}{\giga\hertz} width, according to the ITU-T standard \cite{standard}) and groups of contiguous slices are adaptively assigned to different traffic requests---forming the so-called \textit{superchannels}---according to their volume, the optical bandwidth of the transceivers in use and the adopted modulation format.
Adjacent superchannels are separated by guardbands, constituted by one or multiple contiguous slices that are left unused. The advantages of flexi-grid networks have been quantified in terms of spectrum utilization reductions up to \SI{30}{\percent} with respect to traditional Wavelength Division Multiplexing (WDM) systems \cite{jinno2009spectrum}.
However, the flexi-grid approach requires more advanced and costly optical devices such as ROADMs with colourless/directionless/contentionless add/drop functionalities equipped with dedicated tunable filters supporting coherent detection \cite{feuer2012advanced}.
Moreover, the flexible spectrum allocation techniques enabled by flexi-grid networks typically induce spectrum fragmentation, as often groups of contiguous slices are formed that cannot be assigned to incoming traffic flows because the superchannel they would form is too narrow.
This issue is further increased by the spectrum continuity constraint, which requires that the same spectrum portion is allocated to a traffic flow along all the physical links it traverses.
Therefore, several studies have identified fragmentation-aware spectrum allocation policies and assessed their performance by determining the blocking probabilities for traffic requests of different sizes \cite{yin2013fragmentation}.
Analytical models based on Markov Chains (MCs) have been proposed to determine these blocking probabilities \cite{yu2013first}.
However, unfortunately, for realistic scenarios these models require unaffordable computational resources due to the high number of states that are required to correctly capture the degrees of freedom offered by the flexible grid, thus introducing scalability limitations.

In order to alleviate the spectrum fragmentation issue and limit the costs of equipment installation without renouncing the benefits of flexi-grid networks in terms of spectrum occupation reduction, alternative semi-flexible approaches have been proposed.
One approach is to group traffic requests according to the number of slices required for transmission, and to place requests belonging to the same group along a dedicated fixed grid, with one edge of their superchannel anchored at a specific frequency \cite{shen2014novel}.
Alternatively, a small set of predefined superchannel widths is defined.
Traffic flows are then allocated in the smallest superchannel they fit in, at the price of leaving some spectrum slices unused \cite{comellas2015worthiness}.
Such scenarios allow for more scalable MC models, as for example is the case in \cite{kim2015analytical} and \cite{kim2016two}.
More specifically, \citet{kim2015analytical,kim2016two} use an approximate MC to obtain blocking probabilities in a two-service semi-flexible optical link with a random spectrum allocation policy under the assumption that traffic demands are categorised in two types according to their bit-rate---high or low, respectively.

In this contribution, we first introduce exact MC models for three spectrum allocation policies.
These models have a state space that grows exponentially with the dimensions of the system, making their use infeasible for realistic scenarios.
Therefore, we also provide reduced-state MC models for the same allocation policies.
The downside of reducing the number of states is that the rates of some of the transitions cannot be precisely determined any more; instead, one can only determine lower and upper bounds.
We compare two approaches to dealing with this indeterminacy.
The first approach simply uses a precise but approximate value for each of the indeterminate transition rates.
This approach was used in \citet{kim2015analytical} for one of the policies, and we construct similar models for the two other allocation policies we consider.
For the second approach, we make use of imprecise MC models to implicitly take into account the partially specified character of the transition rates.

Similar approaches were already put forward by us in \cite{2017Rottondi}, where imprecise MC models were applied to optical networks for the first time.
The present contribution---an extended version of \cite{submitted_to_IEEE}, a paper that is currently under peer-review in IEEE Transactions on Communications---extends \cite{2017Rottondi}.
Whereas in \cite{2017Rottondi} we only considered a policy-independent imprecise MC model, we here also construct an imprecise MC model for each of the three policies.
Furthermore, we provide a more thorough explanation on how to compute lower and upper bounds on the blocking probabilities.
Finally, we illustrate the benefits of our imprecise MC models over the exact or approximate MC models by means of extensive numerical experiments.

The remainder of this contribution is organised as follows.
We start in Section~\ref{sec:related} by presenting an overview of the related scientific literature.
Next, we provide some basic background on imprecise probabilities and imprecise continuous-time MCs in Section~\ref{sec:MCIntroduction} and introduce the two-service optical link scenario under study in Section~\ref{ssec:MCMod:OpticalLink}.
We then go on in Section~\ref{sec:MCModels} to construct two precise MC models---an exact and an approximate one---for each of the three considered allocation policies.
Section~\ref{sec:iMCModels} presents our main contribution, which is to introduce four imprecise MC models---one for each of the three allocation policies and a common policy-independent one.
Finally, Section~\ref{sec:Results} numerically compares the scalability of our (exact and approximate) precise MC models and our (policy-dependent and policy-independent) imprecise MC models.
We conclude this contribution in Section~\ref{sec:conclusions}.
For the sake of clarity, all proofs have been relegated to the appendix.

	\section{Related Work}
\label{sec:related}
A rich body of research work on flexi-grid optical networks has appeared in the last few years.
The reader is referred to \cite{zhang2013survey} and \cite{tomkos2012survey} for a thorough overview.
More specifically, both the static and the dynamic Routing and Spectrum Assignment (RSA) problem in flexi-grid networks has been extensively addressed; for a survey of the most relevant literature we refer to~\cite{azodolmolky2009survey}.
In the present contribution, we focus on a single-link dynamic scenario.
The arrival and departure processes of traffic requests are random, and requests are either accommodated in a spectrum portion of the link or rejected in real-time.

Different semi-flexible approaches have been proposed to mitigate the issue of spectrum fragmentation.
\Citet{comellas2015worthiness} propose to partition the flexible grid in blocks of a fixed number of slices and to assign one block to each traffic request, regardless of its size, possibly leaving some unused slices within the block.
If a single block is not sufficient to accommodate the whole traffic demand, multiple contiguous blocks are assigned to it.
Alternatively, \citet{6855763} propose to reserve a dedicated spectrum portion to high bit-rate signals, whereas in \cite{shen2014novel} each specific bit-rate signal uses its own dedicated fixed grid and starts from predefined anchor frequencies.
In this contribution, we adopt the latter approach.

Several authors have investigated the unfairness of blocking probabilities in semi-flexible optical networks in the two-service scenario.
\citet{kim2015analytical} propose an approximate MC model for the calculation of blocking probabilities of the two types of traffic requests along a single optical link, assuming that spectrum allocation is performed randomly.
This model has been refined in \cite{kim2015blocking,wang2014,kim2016two} to include spectrum portions exclusively reserved for each of the two connection types.
In this contribution, we generalise the original model by means of an imprecise MC, which is an MC that allows for imprecise---partially specified---transition rates.
In this way, we can provide guaranteed lower and upper bounds on the performance of a wide class of spectrum allocation policies.
As exemplificative cases, we consider three such cases: the random, least-filled and most-filled allocation policies.
Furthermore, we also use our partially specified transition rates to provide guaranteed lower and upper bounds on the performance of \textit{any} spectrum allocation policy.

A general introduction to the theory of imprecise probabilities can be found in \cite{2014ITIP}.
This theory has been used to generalise both discrete-time and continuous-time MCs, the latter more recently than the former.
A basic treatment of imprecise continuous-time MCs can be found in \cite{2015Skulj,2016Krak,2015Troffaes,2017Erreygers,2016DeBock}.
For their discrete-time counterpart, see for example \cite{2013Skulj,2016Lopatatzidis}.

	\section{Background}
\label{sec:MCIntroduction}

	\subsection{Imprecise Probabilities}
Whenever it is impossible or impractical to provide, obtain or compute exact values for the probability of some event, or the expectation of some function, the theory of imprecise probabilities allows for these quantities to be described `imprecisely', using lower and upper bounds.
For a detailed treatment of this theory, the interested reader is referred to \cite{Walley91} and \cite{2014ITIP}.
For our present purposes, and in the context of stochastic processes, it suffices to understand the following basic concepts.

In the remainder of this paper, we will use $(X_t)_{t \in \mathbb{R}_{\geq0}}$\footnote{%
	$\mathbb{R}$ denotes the set of real numbers, while the set of all non-negative real numbers is denoted by $\mathbb{R}_{\geq 0}$.
	Furthermore, we use $\mathbb{N}$ to denote the set of all natural numbers (including zero), and $\mathbb{N}_{> 0}$ to denote the set of all strictly positive natural numbers.
} to denote a generic continuous-time stochastic process, where for all $t \in \mathbb{R}_{\geq 0}$ the state $X_t$ is a random variable that takes values $x$ in a finite state space $\mathscr{X}$.
Moreover, we let $\mathbb{I}_{A}$ be the indicator $\mathbb{I}_{A} \colon \mathscr{X} \to \{0, 1 \}$  of the event $A \subseteq \mathscr{X}$, defined by $\mathbb{I}_{A}(x)\coloneqq1$ if $x\in A$ and $\mathbb{I}_{A}(x)\coloneqq0$ otherwise.

For our present purposes, the most important notion is that of a conditional lower expectation, which can be interpreted as a lower bound on a conditional expectation.
More formally, for any $s,t \in \mathbb{R}_{\geq 0}$ such that $t\geq s$, any $x_s \in \mathscr{X}$ and any function $f \colon \mathscr{X} \to \mathbb{R}$, the lower expectation of $f$ at time $t$, conditional on $X_s=x_s$, is
\[
	\underline{\mathrm{E}}(f(X_t) \vert X_s = x_s)
	\coloneqq\inf\limits_{\mathrm{E} \in \mathscr{E}} \mathrm{E}(f(X_t) \vert X_s = x_s),
\]
where $\mathscr{E}$ is the set of conditional expectations that corresponds to some set of stochastic processes $\mathbb{P}$---see Section~\ref{subsec:impreciseMCs} for a concrete example of such a set of processes $\mathbb{P}$.

The reason why we can focus on lower expectations, and not on upper expectations or lower/upper probabilities, is because the latter can all be obtained as special cases.
On the one hand, upper expectations are conjugate to lower expectations, in the sense that
\begin{equation}\label{eq:upperEconjugate}
	\overline{\mathrm{E}}(f(X_t) \vert X_s = x_s)
	= - \underline{\mathrm{E}}(- f(X_t) \vert X_s = x_s).
\end{equation}
On the other hand, lower and upper probabilities are special cases of lower and upper expectations, in the sense that for any event $A \subseteq \mathscr{X}$:
\begin{align}
\label{eqn:DefinitionOfLowerProbability}
	\underline{\mathrm{P}}(X_t \in A \vert X_s = x_s)
	&= \underline{\mathrm{E}}(\mathbb{I}_{A}(X_t) \vert X_s = x_s)
\intertext{and}
\label{eqn:DefinitionOfUpperProbability}
	\overline{\mathrm{P}}(X_t \in A \vert X_s = x_s)
	&= \overline{\mathrm{E}}(\mathbb{I}_{A}(X_t) \vert X_s = x_s).
\end{align}
Lower and upper probabilities are also conjugate, in the sense that
\[
	\overline{\mathrm{P}}(X_t \in A \vert X_s = x_s)= 1 - \underline{\mathrm{P}}(X_t \not\in A \vert X_s = x_s).
\]

	\subsection{Imprecise Continuous-Time Markov Chains}
\label{subsec:impreciseMCs}
Recently, several authors have used the theory of imprecise probabilities to develop the notion of an imprecise continuous-time Markov chain, which we will here abbreviate as `imprecise MC'~\cite{2015Skulj,2016DeBock,2016Krak,2017Krak,2017Erreygers}.

The starting point are traditional precise (homogeneous) MCs, which are characterised by a transition rate matrix $Q$ (a real-valued square matrix whose off-diagonal elements are non-negative and whose rows sum to zero).
For any $x,y \in \mathscr{X}$ and any $s, t \in \mathbb{R}_{\geq 0}$ such that $s \leq t$, the $(x,y)$-component of the transition matrix $T_{s}^{t}$ is then
\[
	T_{s}^{t}(x, y)
	\coloneqq \mathrm{P}(X_{t} = y \vert X_{s} = x)
	= [e^{(t - s) Q}](x,y),
\]
where $e^{(t-s) Q}$ is the matrix exponential of $(t - s) Q$, defined as
\begin{equation}
\label{eqn:Matrix exponential}
	e^{(t-s) Q}
	\coloneqq \lim_{n \to +\infty} \sum_{k=0}^{n} \frac{(t-s)^k}{k!} Q^k
	= \lim_{n \to +\infty} \left( I + \frac{t-s}{n} Q \right)^n.
\end{equation}

Instead of a single precisely specified transition rate matrix $Q$, an imprecise MC now considers a (non-empty and bounded) set $\mathscr{Q}$ of transition rate matrices.
In practice, this is typically useful in cases where the values of the transition rates $Q(x,y)$ cannot be determined exactly, as is the case in Section~\ref{ssec:ApprMCMod:TransitionRates}.

More formally, instead of a single homogeneous MC, an imprecise MC considers the set $\mathbb{P}_{\mathscr{Q}}$ of all MCs that are consistent with $\mathscr{Q}$, in the sense that at every point in time $t \in \mathbb{R}_{\geq 0}$, and for $\Delta\in\mathbb{R}_{\geq0}$ sufficiently small, the transition matrix $T_t^{t+\Delta}$ is approximately equal to $I+\Delta Q_t$, for some $Q_t\in\mathscr{Q}$.
Note that the Markov chains in this set $\mathbb{P}_{\mathscr{Q}}$ are not assumed to be homogeneous, in the sense that $Q_t$ is not required to be constant. The only thing that is assumed about $Q_t$ is that it is a---possibly unknown---function of time that takes values in $\mathscr{Q}$.

Since we consider a set of MCs, the transition matrices $T_s^t$ are no longer uniquely known---as was the case for precise MCs.
Instead, an imprecise MC is characterised by a lower transition operator $\underline{T}_{s}^t$.
For any $f \colon \mathscr{X} \to \mathbb{R}$ and any $s,t \in \mathbb{R}_{\geq 0}$ such that $t\geq s$, its value $\underline{T}_{s}^t f$ is again a real-valued function on $\mathscr{X}$, defined by
\begin{equation}\label{eq:defoflowerT}
	[\underline{T}_{s}^t f](x_s)
	\coloneqq \underline{\mathrm{E}}(f(X_t) \vert X_s = x_s)~\text{for all $x_s\in\mathscr{X}$,}
\end{equation}
where $\underline{\mathrm{E}}(f(X_t) \vert X_s = x_s)$ is the infimum of the conditional expectations that are induced by the set of consistent processes $\mathbb{P}_{\mathscr{Q}}$.
Of course, determining the set of all consistent processes $\mathbb{P}_{\mathscr{Q}}$ explicitly and then computing the infimum of the corresponding expectations is infeasible, if not impossible.
Fortunately, this is not always necessary because the lower transition operator $\underline{T}_s^t$ can often also be characterised by a non-linear version of the matrix exponential \eqref{eqn:Matrix exponential}:
\begin{equation}
\label{eqn:LimitComputationOfTheLowerTransitionOperator}
	\underline{T}_s^t
	= \lim_{n \to +\infty} \left( I + \frac{t-s}{n} \underline{Q} \right)^n,
\end{equation}
where $\underline{Q}$ is the so-called \emph{lower transition rate operator} of $\mathscr{Q}$, which maps any $f \colon \mathscr{X} \to \mathbb{R}$ to $\underline{Q}f \colon \mathscr{X} \to \mathbb{R}$, defined by
\begin{equation}\label{eq:Qlower}
	[\underline{Q} f](x)
	\coloneqq \inf \left\{ [Q f](x) \colon Q \in \mathscr{Q} \right\}~\text{for all $x\in\mathscr{X}$.}
\end{equation}
As proved in \cite{2016DeBock}, a sufficient condition for this to be possible is that $\mathscr{Q}$ has separately specified rows, which more or less means that for every $f\colon\mathscr{X}\to\mathbb{R}$, there is some $Q\in\mathscr{Q}$ such that $[\underline{Q}f](x)=[Qf](x)$ for all $x\in\mathscr{X}$; see \cite[Definition~7.3]{2016Krak} for a formal definition.
A lower transition rate operator $\underline{Q}$ also has a norm, which is equal to
\[
	\Vert \underline{Q} \Vert
	\coloneqq \max \{ \vert [\underline{Q} \mathbb{I}_{x}](x) \vert \colon x \in \mathscr{X} \}.
\]

That said, in this contribution, we will be solely interested in providing lower and upper bounds on the value of $\mathrm{P}(X_t \in A \vert X_0 = x_0)$ as $t$ approaches infinity, that is, we are interested in
$
	\lim\limits_{t \to +\infty} \underline{\mathrm{P}}(X_t \in A \vert X_0 = x_0)
$
and
$
	\lim\limits_{t \to +\infty} \overline{\mathrm{P}}(X_t \in A \vert X_0 = x_0).
$
As shown in \cite{2016DeBock}, these limits always exist.
If they furthermore do not depend on $x_0 \in \mathscr{X}$, then we call them the \emph{lower and upper limit probability} of $A$, and denote them by $\underline{\pi}_{A}$ and $\overline{\pi}_{A}$, respectively.
A sufficient condition for the existence of $\underline{\pi}_{A}$ and $\overline{\pi}_{A}$ is the ergodicity of the lower transition rate operator $\underline{Q}$, in the sense that for any real-valued function $f$ on $\mathscr{X}$ and any $s \in \mathbb{R}_{\geq 0}$, $\lim_{t \to +\infty} [\underline{T}_{s}^{t} f](x_{s})$ does not depend on the initial state $x_{s} \in \mathscr{X}$ \cite{2016DeBock}.
A simple necessary and sufficient condition for ergodicity is provided in \cite[Theorem~19]{2016DeBock}.

Provided that $\mathscr{Q}$ has separately specified rows and $\underline{Q}$ is ergodic, it follows from \eqref{eq:upperEconjugate}--\eqref{eqn:DefinitionOfUpperProbability}, \eqref{eq:defoflowerT} and \eqref{eqn:LimitComputationOfTheLowerTransitionOperator} that
\begin{align}
\label{eqn:Limit lower probability}
	\underline{\pi}_{A}
	&= \lim_{t \to +\infty}\lim_{n \to +\infty}\left[ \left( I + \frac{t}{n} \underline{Q} \right)^n \mathbb{I}_{A} \right](x_0)
\intertext{and}
\label{eqn:Limit upper probability}
	\overline{\pi}_{A}
	&= -\lim_{t \to +\infty}\lim_{n \to +\infty}\left[ \left( I + \frac{t}{n} \underline{Q} \right)^n(-\mathbb{I}_{A})\right](x_0)
\end{align}
for any $x_{0} \in \mathscr{X}$, which allows us to compute them both by choosing $t$ and $n$ sufficiently large.
Unlike for the case of precise MCs, where literature offers a plethora of methods to compute limit probabilities, this is---to the best of our knowledge---currently the only method to compute lower or upper limit probabilities for imprecise MCs.

	\subsection{Numerical Approximation Method}
\label{ssec:iMC:ApproximationMethod}
Assume we have some ergodic lower transition rate operator $\underline{Q}$ and some event $A \subseteq \mathscr{X}$.
We can then use Alg.~\ref{alg:Standard}---a slightly modified version of \cite[Algorithm~1]{2016Krak}---to compute the lower limit probability $\underline{\pi}_{A}$ of the event $A \subseteq \mathscr{X}$.
This algorithm uses the variation norm, defined for all $f \colon \mathscr{X} \to \mathbb{R}$ as
$
	\Vert f \Vert_{v}
	\coloneqq (\max f - \min f) / 2,
$
and the midpoint, defined as
$
	\midpoint(f)
	\coloneqq ( \max f + \min f) / 2.
$
Note that to compute $\overline{\pi}_{A}$, we simply need to change line~\ref{line:Standard:First} in Alg.~\ref{alg:Standard} to ``$g_{0} \gets - \mathbb{I}_{A}$'' and line~\ref{line:Standard:End} to ``$\hat{\pi}_{A} \gets \midpoint(-g_{n})$''.

	\begin{algorithm}[t]
		\caption{Approximating the lower limit probability \label{alg:Standard}}
		\DontPrintSemicolon
		\KwData{An ergodic lower transition rate operator $\underline{Q}$, an event $A \subseteq \mathscr{X}$, a time step $\delta \in (0, 2 / \Vert \underline{Q} \Vert)$, a relative tolerance $\phi \in \mathbb{R}_{> 0}$ and a maximum number of iterations $N \in \mathbb{N}$.}
		\KwResult{An approximation $\hat{\pi}_{A}  = \underline{\pi}_{A} \pm \epsilon'$}

		$i \gets 0$ \;
		$g_{0} \gets \mathbb{I}_{A}$\; \nllabel{line:Standard:First}
		\While{$\Vert g_{i} \Vert_{v} > \phi \vert \midpoint(g_{i}) \vert$ \And $i < N$}{
			$
			i \gets i+1$ \;
			$g_{i} \gets g_{i-1} + \delta \underline{Q} g_{i-1}$ \;
		}
		$\hat{\pi}_{A} \gets \midpoint(g_{i})$ \; \nllabel{line:Standard:End}
		\Return $\hat{\pi}_{A}$
	\end{algorithm}

It is worth noting that we need to specify a time step $\delta \in (0, 2 / \Vert \underline{Q} \Vert)$, a maximal relative tolerance $\phi \in \mathbb{R}_{> 0}$ and a maximum number of iterations $N \in \mathbb{N}$.

From \cite[Proposition~11]{2017Erreygers} and the theory of ergodic imprecise discrete-time Markov chains \cite{2013Skulj}, it follows that if $\underline{Q}$ is ergodic, then for any $\delta \in (0, 2 / \Vert \underline{Q} \Vert)$,
$
	\lim_{n \to +\infty} [(I + \delta \underline{Q})^{n} \mathbb{I}_{A}](x_{0})
$
is the same for all $x_{0} \in \mathscr{X}$.
However, this limit value is not necessarily the same for all allowable values of $\delta$.
Empirically, we observe that for the lower transition rate operators we introduce in Sections~\ref{ssec:iMCMod:PolicyDependent} and \ref{ssec:iMCMod:PolicyIndependent} this limit value increases (i.e., the obtained lower bound is tighter) for decreasing $\delta$.
One way to empirically confirm that the lower bound $\hat{\pi}_{A}$ obtained using Alg.~\ref{alg:Standard} is sufficiently tight is proposed---albeit in a slightly different form---in \cite{2015Troffaes}: run Alg.~\ref{alg:Standard} for the step sizes $\delta$ and $\delta / 2$, and compare the obtained bounds $\hat{\pi}_{A,1}$ and $\hat{\pi}_{A,2}$.
If $\vert \hat{\pi}_{A,1} - \hat{\pi}_{A,2} \vert < \phi \hat{\pi}_{A,2}$---for example, for $\phi = 10^{-m}$, if $\hat{\pi}_{A,1}$ and $\hat{\pi}_{A,2}$ differ only after the $m$-th significant digit---then we can be relatively confident that the obtained bound is sufficiently tight.

If one wants a theoretically guaranteed instead of empirically verified measure of the accuracy of the bound, things get a little more intricate.
To the best of our knowledge, existing methods are based on the absolute error
$
	\epsilon_{a}
	\coloneqq \vert \underline{\pi}_{A} - \midpoint g_{n} \vert.
$
Let $g_{0}, g_{1}, \dots, g_{n}$ be obtained by running Alg.~\ref{alg:Standard}.
Then by \cite[Theorem~5 and Proposition~12]{2017Erreygers} a theoretically guaranteed upper bound for the absolute error $\epsilon_{a}$ is
$
	\epsilon'
	\coloneqq \max \left\{ 2 \delta^2 \Vert \underline{Q} \Vert^{2} \sum_{i=0}^{n-1} \Vert g_{i} \Vert_{v}, \Vert g_{n} \Vert_{v} \right\}.
$
However, as illustrated in \cite[Example~6]{2017Erreygers}, this upper bound can be overly conservative.
Therefore, and because in the current setting it makes more sense to consider the relative error instead of the absolute error, we would here opt to use the empirical way of assessing the accuracy.
Nevertheless, in Section~\ref{ssec:Numerical assessment:lower and upper blocking probabilities} we simply use $\delta = \nicefrac{2}{(4 \Vert \underline{Q} \Vert)}$, which---although we do not explicitly mention it there---turns out to yield the required accuracy in all cases, except for two cases where the computations preemptively stop because the maximum number of iterations is reached.

	\section{The Single Optical Link Under Study}
\label{ssec:MCMod:OpticalLink}

We consider a single optical-fiber link with overall spectrum availability $T$.
The spectrum is partitioned in slices of width $F$, for a total number of $S \coloneqq T/F$ slices.
For the sake of convenience, we assume that $T$ is an integer multiple of $F$.
Slices are sequentially numbered from $1$ to $S$.

The optical link is used to transmit two types of flows with distinct bandwidth demands: flows of type 1 require $n_1 > 0$ slices, whereas flows of type 2 require $n_2 n_1 > n_1$ slices.
Type 1 (respectively type 2) flows arrive according to a Poisson process with arrival rate $\lambda_1$ (respectively $\lambda_2$) and, if assigned to some free slices in the spectrum, cease after a holding time that is exponentially distributed and has average service time $1/\mu_1$ (respectively $1/\mu_2$).
We are not interested in edge cases, so throughout this contribution we assume that $\lambda_1$, $\lambda_2$, $\mu_{1}$ and $\mu_{2}$ are all known positive real numbers.

\begin{figure}[ht]
	\centering
	\begin{tikzpicture}[%
			scale=0.9,%
			every node/.style={scale=0.9},%
			decoration = brace%
		]%
		\clip (-.1, -.15) rectangle (9.2, 1.7);
		\draw[->] (0,0) -- (9.2,0);
		\foreach \x in {0, 1.5, 3, 4.5, 6, 7.5} {
			\draw (\x,0) -- (\x, 1);
			\foreach \Dx in {0.5, 1} {
				\draw[dashed] (\x+\Dx,0) -- (\x+\Dx,1);
			}
			\foreach \dx in {.25, .75, 1.25} {
					\draw[dotted] (\x+\dx,0) -- (\x+\dx,1);
			}
		}
		\draw (9,0) -- (9,1);
		\draw[decorate, line width=.5 pt] (-.05,1.2) -- node[auto=left] {$F$} (.255,1.2);
		\draw[decorate, line width=.5 pt] (1.5,1.2) -- node[auto=left] {$n_1 F$} (2,1.2);
		\draw[decorate, line width=.5 pt] (3,1.2) -- node[auto=left] {$n_2 n_1 F$} (4.5,1.2);
		\begin{scope}
			\clip (0,0) rectangle (10, 5);
			\foreach \x in {0, 4.5, 6.5} {
				\draw[blue,fill=blue, opacity=.5] (\x+.75, 0) ellipse (.70 and .75);
			}
			\foreach \x in {1.5, 2, 2.75, 4, 6, 8} {
				\draw[orange,fill=orange, opacity=.5] (\x+.25,0) ellipse (.20 and .75);
			}
		\end{scope}
		\draw[line width = 2 pt, red] (2.75, -.1) -- (3.25, .7);
		\draw[line width = 2 pt, red] (3.25, -.1) -- (2.75, .7);
		\draw[line width = 2 pt, red] (6.5, -.1) -- (8, .7);
		\draw[line width = 2 pt, red] (8, -.1) -- (6.5, .7);
	\end{tikzpicture}
	\caption{Example of the (super)channel ordering in a single optical link with $n_1=2$, $n_2=3$ and $S=36$.}
	\label{fig:modes}
\end{figure}
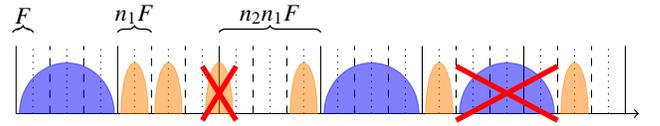
As depicted in Fig.~\ref{fig:modes}, we consider two overlapping grids of different granularity.
The first consists of a sequence of adjacent channels of width $n_1 F$: the first channel comprises slices $1,\dots,n_1$, the second comprises slices $n_1+1,\dots,2n_1$, and so on.
Similarly, the second grid consists of a sequence of adjacent superchannels of width $n_2 n_1 F$.
The number of channels of size $n_1 F$ is thus $m_1 \coloneqq S/n_1$, whereas the number of superchannels of size $n_2 n_1 F$ is $m_2 \coloneqq m_1/n_2$.
For the sake of convenience, we assume that $S$ is an integer multiple of $n_1 n_2$.

An incoming type 1 flow, in the remainder referred to as a \emph{type 1 request}, must be assigned to a free channel of width $n_1 F$, whereas an incoming type 2 flow, in the remainder referred to as a \emph{type 2 request}, must be assigned to a free superchannel of width $n_2 n_1 F$.
If there is only a single free (super)channel, we always assign the request to this free (super)channel.
If there are multiple free (super)channels, we assign the request to one of the free (super)channels according to some spectrum allocation policy.
For type 2 requests, it will become evident in the remainder that the allocation policy is of no importance to our analysis.
For type 1 requests, we restrict ourselves to allocation policies that only depend on the number of type 1 flows currently occupying each superchannel and not on the specific ordering of the type 1 flows along the spectrum.

We consider three such allocation policies as exemplificative cases.
The Random Allocation (RA) policy assigns a type 1 request to a randomly selected free channel if possible, where every free channel has the same probability of being selected \cite{kim2015analytical,kim2016two}.
Alternatively, we can assign the type 1 request to one of the free channels in a partially occupied (i.e., non-empty and non-full) superchannel that contains either the lowest number of type 1 flows or the highest number of type 1 flows.
The former is called the Least-Filled (LF) policy, while the latter is called the Most-Filled (MF) policy.
For both of these policies, if all superchannels are empty or full, the type 1 request is assigned to one of the channels in an empty superchannel---provided they are not all full of course.
Throughout the remainder, we use AP to denote a generic allocation policy.

If no (super)channel of the required size is available, the incoming traffic request is \emph{blocked}.
Determining the blocking probability of type 1 or type 2 traffic requests (or more specifically, the long-term limit of the fraction of incoming requests of type 1 or type 2 that are blocked), denoted by BP1 and BP2 respectively, is the main objective of this contribution.

	\section{Precise Markov Chain Models}
\label{sec:MCModels}

	\subsection{Exact Model}
\label{ssec:MC Models:Exact model}

\subsubsection{Detailed State Space Description}
To determine the blocking probabilities of type 1 and type 2 requests, we will construct policy-dependent MC models and determine the limit probabilities of the event that no type 1 (or type 2) flow can be allowed into the system.\footnote{%
	Note that although the limit probability has a different definition than the blocking probability, it is the customary way to determine the blocking probability as it follows from the well-known PASTA property and the ergodic theorem for (irreducible) Markov chains that these are in fact (almost surely) equal.
}
The state description of these MC models should allow us to (i) determine whether or not an incoming flow is blocked or allowed into the system, and (ii) accurately model the allocation of the flows and the completion of their holding times.
Due to the memorylessness of the exponential distribution and our requirement that the allocation policy for type 1 requests only depends on the number of type 1 flows currently occupying each superchannel, a sufficiently detailed state description is $(i_0,i_1,\dots,i_{n_2})$.
In this tuple, $i_k$ counts the number of superchannels occupied with $k$ type 1 flows and no type 2 flows.
Let $I$ be defined as $I \coloneqq \sum_{k=0}^{n_2}i_k$, then $m_2-I$ is the number of superchannels that are occupied by a type 2 flow.
To ensure feasibility, we require that $I \leq m_2$ and that $i_k \geq 0$ for all $k$ in $\{0, \dots, n_2\}$, which yields the detailed state space
$
	\mathscr{X}_{\mathrm{det}}
	\coloneqq \left\{ (i_{0}, \dots, i_{n_{2}}) \in \mathbb{N}^{n_{2}+1} \colon \sum_{k = 0}^{n_{2}} i_{k} \leq m_{2}  \right\}.
$
The total number of states exhibits an $\mathcal{O}(m_{2}^{n_{2}}) = \mathcal{O}((m_1 / n_2)^{n_{2}})$ dependency on the total number of channels $m_1$ and on the number of channels contained in a superchannel $n_2$.

\subsubsection{Policy-Dependent Transition Rates}

Note that $i_0$ counts the number of empty superchannels, whereas $R \coloneqq \sum_{k=0}^{n_2-1}i_k(n_2-k)$ is the total number of free channels.
\begin{figure}[ht]
	\centering
	\begin{tikzpicture}[%
			scale=0.8,
			every node/.style={scale=0.8},
			every state/.style={rounded rectangle}
		]
		\node[state, fill=gray!20] (cent) {$i_0, \dots, i_k, \dots, i_{n_2}$};
		\node[state, above left = .8 cm and 0.7 cm of cent] (NW) {$i_0+1, \dots, i_k, \dots, i_{n_2}$};
		\node[state, above right = .8 cm and 0 cm of cent] (NE) {$i_0, \dots, i_k-1, i_{k+1}+1, \dots, i_{n_2}$};
		\node[state, below right = .8 cm and 0.7 cm of cent] (SE) {$i_0-1, \dots, i_k, \dots, i_{n_2}$};
		\node[state, below left = .8 cm and 0 cm of cent] (SW) {$i_0, \dots, i_{k-1}+1, i_{k}-1, \dots, i_{n_2}$};
		\draw[->]
			(cent) edge[bend right, auto=left]
				node[pos=.35, anchor=east] {$(m_2 - I) \mu_2$}
			(NW)
			(cent) edge[bend right, auto=right]
				node[pos=.6, anchor=west] {$k i_k \mu_1$}
			(SW)
			(cent) edge[bend right, auto=left]
				node[pos=.5, anchor=east] {$\lambda_{\mathrm{AP}}$}
			(NE)
			(cent) edge[bend right, auto=right]
				node[pos=.4, anchor=west] {$\lambda_2$}
			 (SE);
		\node[below = 0pt of NW, xshift=-8pt, font=\footnotesize] {(if $I < m_{2}$)};
		\node[below = 0pt of NE, xshift=25pt, font=\footnotesize] {(if $R > 0$, $i_{k} > 0$)};
		\node[above = 0pt of SE, xshift=-5pt, font=\footnotesize] {(if $i_0>0$)};
		\node[above = 0pt of SW, xshift=-15pt, font=\footnotesize] {(if $i_{k} > 0$)};
	\end{tikzpicture}
	\caption{State transition diagram of the proposed precise and exact Markov chain}
	\label{fig:ExactStateSpace}
\end{figure}
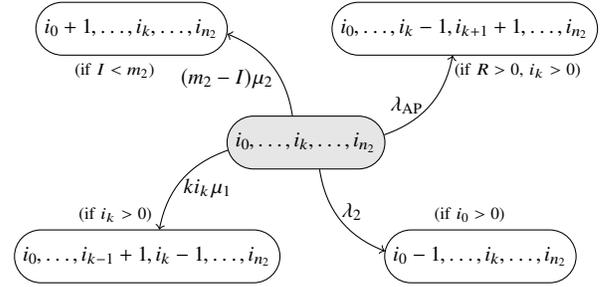
As depicted in Fig.~\ref{fig:ExactStateSpace}, the following transitions can occur (i.e., have positive rate).
If a new type 2 flow arrives and there is a free superchannel (i.e., $i_0 > 0$), the transition $(i_0,i_1,\dots,i_{n_2}) \rightarrow (i_0-1,i_1,\dots,i_{n_2})$ takes place.
As type 2 flows arrive according to a Poisson process with rate $\lambda_{2}$, this transition occurs with the same rate.
Conversely, when the holding time of a type 2 flow expires, the following transition occurs: $(i_0,i_1,\dots,i_{n_2})\rightarrow(i_0+1,i_1,\dots,i_{n_2})$.
There are $(m_2 - I)$ allocated type 2 flows, and the holding time of each of these flows is (independently) exponentially distributed with rate $\mu_{2}$.
Hence, it follows from the properties of the exponential distribution that this transition occurs with rate $(m_2-I)\mu_2$.
Similarly, the expiration of the holding time of a type 1 flow corresponds to the state transition $(i_0,\dots,i_k,\dots, i_{n_2}) \rightarrow (i_0,\dots,i_{k-1}+1,i_{k}-1, \dots, i_{n_2})$ which, for all $k$ in $[1, n_2]$, has rate $k i_k \mu_1$.

The transition that corresponds to the assignment of a type 1 request, which can only occur if there is a free channel (i.e., $R > 0$), depends on the adopted spectrum allocation policy AP.
If we adopt the RA policy, then the state transition $(i_0,\dots,i_k,\dots, i_{n_2}) \rightarrow (i_0,\dots,i_k-1,i_{k+1}+1, \dots, i_{n_2})$ occurs for all $k$ in $[0, n_2-1]$ with rate $\lambda_{\mathrm{RA}}= (\nicefrac{1}{R}) \lambda_1 i_k (n_2-k)$.
Alternatively, if we adopt the LF or MF policies, the state transition $(i_0,\dots,i_{k_{\mathrm{AP}}},\dots, i_{n_2}) \rightarrow (i_0,\dots,i_{k_{\mathrm{AP}}}-1,i_{k_{\mathrm{AP}}+1}+1, \dots, i_{n_2})$ occurs with rate $\lambda_{\mathrm{LF}}=\lambda_{\mathrm{MF}}=\lambda_1$, where $k_{\mathrm{AP}}$ depends on the policy AP.
If there is at least one superchannel that is partially occupied by a type 1 flow (i.e., if there is some $k \in [1,n_2-1]$ such that $i_k>0$), then $k_{\mathrm{LF}}=\min\{k\in[1,n_2-1]\colon i_k>0\}$ and $k_{\mathrm{MF}}=\max\{k\in[1,n_2-1]\colon i_k>0\}$.
Otherwise, that is, if all the superchannels are either completely free or completely occupied (i.e., if $i_k=0$ for all $k \in [1,n_2-1]$), then $k_{\mathrm{LF}}=k_{\mathrm{MF}}=0$.
Note that in case $n_2 = 2$, $k_{\mathrm{LF}} = k_{\mathrm{MF}}$.

\subsubsection{Irreducibility and Limit Probabilities}
For each of the three spectrum allocation policies that we consider, the rates that were specified above determine a unique transition rate matrix.
One can easily verify that these transition rate matrices $Q_{\mathrm{RA}}$, $Q_{\mathrm{LF}}$ and $Q_{\mathrm{MF}}$ are irreducible; see Appendix~\ref{app:PreciseErgodicity} for a formal proof.

Due to this irreducibility, for every event $A\subseteq\mathscr{X}_{\mathrm{det}}$ they each determine a unique limit probability $\pi_A$.
The blocking probability BP1 experienced by type 1 requests is (almost surely) equal to the limit probability of the event $A_1=\{(i_0, \dots, i_{n_2})\in\mathscr{X}_{\mathrm{det}}\colon R=0\}$, whereas the blocking probability BP2 experienced by type 2 requests is (almost surely) equal to the limit probability of the event $A_2=\{(i_0, \dots, i_{n_2})\in\mathscr{X}_{\mathrm{det}}\colon i_0=0\}$.

	\subsection{Reduced State Space Description}
\label{ssec:ApprMCMod:StateSpace}
The main reason that using a detailed Markov model---one with the detailed state space $\mathscr{X}_{\mathrm{det}}$---to determine the blocking probabilities is infeasible for large systems, is that determining the blocking probabilities becomes computationally intensive due to the exponential dependency of the number of states on the system dimensions.
Hence, one approach to reduce the duration of the computations is to reduce the number of states in the MC models.
To that end, we now present alternative policy-dependent precise MC models that adopt the more compact---though less informative---state description that, in case of the random allocation policy, was first introduced by \citet{kim2015analytical}.

\subsubsection{The Reduced State Space}
A coarser state space description that still allows us to determine whether or not a request is blocked, is the triplet $(i,j,e)$.
In this triplet, $0 \leq i \leq m_1$ (respectively $0 \leq j \leq m_2$) counts the number of type 1 (respectively type 2) flows currently allocated, and $0 \leq e \leq m_2$ counts the number of free superchannels.
To ensure feasibility, it must hold that $m_2 \leq i + j + e$ and $i+(j+e)n_2 \leq m_1$.
Note that the first inequality is not (explicitly) mentioned by \citet{kim2015analytical}, but is nevertheless required to ensure that all superchannels are accounted for.
Enforcing these two feasibility constraints yields the reduced state space
\[
	\mathscr{X}_{\mathrm{red}}
	\coloneqq \left\{ (i,j,e) \in \mathbb{N}^{3} \colon m_2 \leq i + j + e, i+(j+e)n_2 \leq m_1 \right\}.
\]

The number of states in this reduced state space $\mathscr{X}_{\mathrm{red}}$ has a $\mathcal{O}(m_1 m_{2}^2) = \mathcal{O}(m_1 (m_1 / n_2)^2)$ dependency on the total number of channels $m_{1}$ and the number of channels that form a superchannel $n_{2}$, which is an improvement over the $\mathcal{O}((m_1 / n_2)^{n_2})$ dependency of the number of states in the detailed state space $\mathscr{X}_{\mathrm{det}}$.
That the reduced state space~$\mathscr{X}_{\mathrm{red}}$ has fewer states than (or at most the same number of states as) the detailed state space~$\mathscr{X}_{\mathrm{det}}$ follows from the fact that the function $\Gamma \colon \mathscr{X}_{\mathrm{det}} \to \mathscr{X}_{\mathrm{red}}$, defined for all $(i_{0}, \dots, i_{n_{2}})$ in $\mathscr{X}_{\mathrm{det}}$ as
\begin{equation}
\label{eqn:ApprMCMod:SurjectiveRelation}
	\Gamma(i_0, \dots, i_{n_2})
	\coloneqq \left( \sum_{k=1}^{n_{2}} k i_{k},\ m_{2} - \sum_{k = 0}^{n_{2}} i_{k},\ i_{0} \right),
\end{equation}
is surjective.
Note that in the special case that $n_2 = 2$, the function $\Gamma$ is invertible:
$
	\Gamma^{-1}(i, j, e)
	= (e, 2 m_2 - i - 2 j - 2 e, i+j+e-m_{2}).
$
Consequently, in that case, there is a one-to-one correspondence between the detailed and reduced state spaces.

\subsubsection{Policy-Dependent Transition Rates}
\label{ssec:ApprMCMod:TransitionRates}
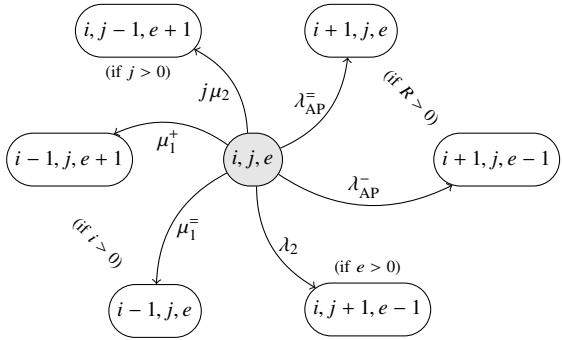
\begin{figure}[ht]
	\centering
	\begin{tikzpicture}[%
			 scale=0.8,
			 every node/.style={scale=0.8},
			every state/.style={rounded rectangle}
		]
		\node[state, fill=gray!20] (cent) {$i, j, e$};
		\node[state, left = 1.2 cm of cent] (W) {$i-1, j, e+1$};
		\node[state, above left = 1 cm and 1 cm of cent] (NW) {$i, j-1, e+1$};
		\node[state, above right = 1 cm and 1 cm of cent] (NE) {$i+1, j, e$};
		\node[state, right = 2 cm of cent] (E) {$i+1, j, e-1$};
		\node[state, below right = 1.28 cm and 1 cm of cent] (SE) {$i, j+1, e-1$};
		\node[state, below left = 1.3 cm and 1 cm of cent] (SW) {$i-1, j, e$};
		\draw[->]
			(cent) edge[bend right, auto=left]
				node[xshift=7pt] {$\mu_1^{+}$}
			(W)
			(cent) edge[bend right, auto=right]
				node [pos=.4, anchor=east] {$j\mu_2$}
			(NW)
			(cent) edge[bend right, auto=right]
				node[pos=.6, anchor=east] {$\lambda^{=}_{\mathrm{AP}}$}
			(NE)
			(cent) edge[bend right, auto=left]
				node[pos=.5, anchor=south] {$\lambda^{-}_{\mathrm{AP}}$}
			(E)
			(cent) edge[bend right, auto=left]
				node[pos=.5, anchor=west] {$\lambda_2$}
			(SE)
			(cent) edge[bend right, auto=right]
				node[pos=.6, anchor=west] {$\mu_1^{=}$}
			(SW);
		\node[below = 0pt of NW, font=\footnotesize] {(if $j > 0$)};
		\node[anchor=base, below right = 1pt and 1pt of NE, font=\footnotesize] {\rotatebox{-45}{(if $R > 0$)}};
		\node[above = 0pt of SE, font=\footnotesize] {(if $e>0$)};
		\node[anchor=base, above left= 2pt and 1pt of SW, font=\footnotesize] {\rotatebox{-45}{(if $i > 0$)}};
	\end{tikzpicture}
	\caption{State transition diagram of the precise but approximate Markov chain and of the proposed imprecise and scalable Markov chain}
	\label{fig:ApproximateStateSpace}
\end{figure}

All non-zero transition rates are schematically depicted in Fig.~\ref{fig:ApproximateStateSpace}.
If a free superchannel is available (i.e., if $e > 0$), allocating a type 2 request corresponds to the transition $(i,j,e) \rightarrow (i,j+1,e-1)$, which therefore occurs with rate $\lambda_2$.
Conversely, the expiration of the holding time of an allocated type 2 flow---which can occur if $j > 0$---corresponds to the transition $(i,j,e) \rightarrow (i,j-1,e+1)$, which therefore occurs with rate $j \mu_2$.

The expiration of the holding time of an allocated type 1 flow---which can occur if $i > 0$---might free up a superchannel or not, depending on whether or not it was the sole type 1 flow in its superchannel.
In order to properly model this behaviour, we need to distinguish three separate cases.
If every superchannel that is neither empty nor occupied by a type 2 flow only contains a single type 1 flow (i.e., if $m_2 = i + j + e$), then the departure of a type 1 request always frees a superchannel.
This corresponds to the transition $(i, j, e) \to (i-1, j, e+1)$, which in this case has rate $\mu_{1}^{+} = i \mu_{1}$.
Note that in this case $(i-1, j, e) \notin \mathscr{X}_{\mathrm{red}}$.
Conversely, if every superchannel that is neither empty nor occupied by a type 2 flow contains at least two type 1 flows (i.e., if $i \geq n_2(m_2-j-e-1)+2$), then the departure of a type 1 flow will never free a superchannel.
This corresponds to the transition $(i, j, e) \to (i-1, j, e)$, which in this case has rate $\mu_{1}^{=} = i \mu_{1}$.
Note that in this case $(i-1, j, e+1) \notin \mathscr{X}_{\mathrm{red}}$.
Unfortunately, in the remaining case that $m_2-j-e < i < n_2(m_2-j-e-1)+2$---or  equivalently, in case $(i-1, j, e+1) \in \mathscr{X}_{\mathrm{red}}$ and $(i-1, j, e) \in \mathscr{X}_{\mathrm{red}}$---the state representation $\mathscr{X}_{\mathrm{red}}$ is in general not sufficiently informative to capture the distribution of the allocated type 1 flows across the superchannels.
In these cases it may therefore not be possible to determine the rate $\mu_1^{=}$ of the transition $(i, j, e) \rightarrow (i-1, j, e)$ and the rate $\mu_1^{+}$ of the transition $(i, j, e) \rightarrow (i-1, j, e+1)$ as a function of $(i,j,e)$.
Consequently, these rates are (possibly unknown) functions of $(i,j,e,t)$, where $t$ is the time of the transition.
All we can say for sure is that they are non-negative, that their sum is equal to $i\mu_1$ and that
\begin{equation}\label{eq:muimprecise}
	\mu_1^{+}(i,j,e,t)
	\in [i_{\min}(i,j,e)\mu_1,i_{\max}(i,j,e)\mu_1],
\end{equation}
where $i_{\min}(i,j,e)\coloneqq\max\{0,2(m_2-j-e)-i\}$ is the minimum number of allocated type 1 flows that are alone in their superchannel and
$
	i_{\max}(i,j,e)
	\coloneqq \left\lfloor \frac{n_2(m_2-j-e)-i}{n_2-1} \right\rfloor
$
is the maximum number of such type 1 flows.
In order to deal with this indeterminacy, the authors of \cite{kim2015analytical} replace the guaranteed bounds \eqref{eq:muimprecise} with a precise and constant estimate $\tilde{\mu}_{1}^{+}(i, j, e)$ for $\mu_1^{+}(i, j, e,. t)$, which is based on the assumption that all the possible situations---that is, all possible distributions of type 1 flows---that are represented by the state $(i,j,e)$ are equally probable.
The approximation error introduced by these estimates will be numerically evaluated in Section~\ref{ssec:Numerical assessment:Accuracy of approximate} further on, by comparing them with the results obtained using the exact model introduced in Section~\ref{ssec:MC Models:Exact model}.
The case $n_2 = 2$ again deserves special attention: we immediately verify that $\mu_{1}^{+}(i, j, e, t)$ is then a constant function of time $t$, as $i_{\min} = i_{\max}$.

If there is an available channel (i.e., if $R\coloneqq m_1-i-jn_2>0$), then a type 1 request can be assigned to a channel in either a non-full superchannel that already contains some type 1 flows or in a completely free superchannel.
We again need to distinguish three cases to properly model this.
If there is no empty superchannel (i.e., if $e = 0$) then, regardless of the spectrum allocation policy, the type 1 request is assigned to a free channel in a non-empty superchannel.
This corresponds to the transition $(i, j, e) \to (i+1, j, e)$, which in this case has rate $\lambda^{=}_{\mathrm{AP}} = \lambda_1$.
Note that in this case $(i+1, j, e-1) \notin \mathscr{X}_{\mathrm{red}}$.
Conversely, if every superchannel that is neither empty nor occupied by a type 2 flow is completely occupied by type 1 flows (i.e., $i = n_2 (m_2 - j - e)$), then a type 1 request is always assigned to an empty superchannel.
This corresponds to the transition $(i, j, e) \to (i+1, j, e-1)$, which in this case has rate $\lambda^{-}_{\mathrm{AP}} = \lambda_1$.
Note that in this case $(i+1, j, e) \notin \mathscr{X}_{\mathrm{red}}$.

In the remaining case, we have that $e > 0$ and $i < n_2 (m_2 - j - e)$.
The two states $(i+1, j, e)$ and $(i+1, j, e-1)$ are then both feasible, and the rates $\lambda^{=}_{\mathrm{AP}}$ and $\lambda^{-}_{\mathrm{AP}}$ of the transitions to these states depend on the used allocation policy.
Regardless of the policy, however, both of these rates should be non-negative and their sum should equal $\lambda_1$.
Hence, we can focus on determining $\lambda^{-}_{\mathrm{AP}}$ because $\lambda^{=}_{\mathrm{AP}} = \lambda_1 - \lambda^{-}_{\mathrm{AP}}$.
For the RA policy, it is shown by \citet{kim2015analytical} that $\lambda^{-}_{\mathrm{RA}}=\lambda_1(\nicefrac{(e n_2)}{(m_1-i-jn_2)})$.
The LF and MF policies only assign a type 1 request to an empty superchannel if all non-empty superchannels containing type 1 requests are completely occupied (i.e., if $i = n_{2} (m_2 - j - e)$).
Consequently, we have that $\lambda^{-}_{\mathrm{LF}} = \lambda^{-}_{\mathrm{MF}} = \lambda_1$ if $i = n_2 (m_2-j-e)$ and $\lambda^{-}_{\mathrm{LF}} = \lambda^{-}_{\mathrm{MF}} = 0$ if $i < n_2 (m_2-j-e)$.

\subsubsection{Irreducibility and Limit Probabilities}
\label{ssec:ApprMCMod:Ergodicity}
For each of the spectrum allocation policies RA, LF and MF, the approximation $\tilde{\mu}_{1}^{+}(i, j, e)$ for $\mu_1^{+}(i, j, e, t)$ in \cite{kim2015analytical} leads to an approximate transition rate matrix $\tilde{Q}_{\mathrm{AP}}$.
Furthermore, since $\lambda_{\mathrm{LF}}^{-} = \lambda_{\mathrm{MF}}^{-}$, we know that $\tilde{Q}_{\mathrm{LF}} = \tilde{Q}_{\mathrm{MF}} \eqqcolon \tilde{Q}_{\mathrm{LM}}$.
As shown in Appendix~\ref{app:PreciseErgodicity}, $\tilde{Q}_{\mathrm{RA}}$ and $\tilde{Q}_{\mathrm{LM}}$ are both irreducible.
Hence, for every event $A\subseteq\mathscr{X}_{\mathrm{red}}$, each of these policy-dependent approximate transition rate matrices leads to a unique limit probability $\pi_A$. 
An approximation for the blocking probability BP1 (respectively BP2) experienced by type 1 (respectively type 2) traffic requests can therefore be obtained by computing the limit probability of the event $A_1=\{(i,j,e)\in\mathscr{X}_{\mathrm{red}} \colon R=0\}$ (respectively $A_2=\{(i,j,e)\in\mathscr{X}_{\mathrm{red}} \colon e=0\}$).

Note that in case $n_2 = 2$, the limit probabilities derived from $\tilde{Q}_{\mathrm{RA}}$ and $\tilde{Q}_{\mathrm{LM}}$ provide exact blocking probabilities rather than approximate ones.
This is a consequence of the one-to-one correspondence between $\mathscr{X}_{\mathrm{det}}$ and $\mathscr{X}_{\mathrm{red}}$ and the equality of $i_{\min}$ and $i_{\max}$, which were both previously mentioned as special cases.

	\section{Reliable and Scalable Imprecise MC Models}
\label{sec:iMCModels}

We now present our main contributions: policy-dependent imprecise MC models that explicitly account for our indeterminacy about the exact value of $\mu_{1}^{+}(i,j,e,t)$ and a common imprecise MC model that is policy-independent.
The former will be used to determine guaranteed lower and upper bounds on the performance of each of our three policies, while the latter yields best and worst-case performance bounds for any policy---so not just the ones that we have studied in detail.

	\subsection{Policy-Dependent Imprecise MC Models}
\label{ssec:iMCMod:PolicyDependent}
For each of the three allocation policies AP, we now consider the set of transition rate matrices $\mathscr{Q}_{\mathrm{AP}}$ that contains all the transition rate matrices that are compatible with the relevant (bounds on the) rates specified in Section~\ref{ssec:ApprMCMod:TransitionRates}.
In particular, for every policy AP and any choice of $\mu_{1}^{+}(i, j, e) \in [i_{\mathrm{min}} \mu_{1}, i_{\mathrm{max}} \mu_{1}]$, we obtain a different matrix $Q_{\mathrm{AP}} \in \mathscr{Q}_{\mathrm{AP}}$
The resulting set~$\mathscr{Q}_{\mathrm{AP}}$ then characterises a lower transition rate operator~$\underline{Q}_{\mathrm{AP}}$ according to \eqref{eq:Qlower}.
In this way, we obtain three sets $\mathscr{Q}_{\mathrm{RA}}$, $\mathscr{Q}_{\mathrm{LF}}$ and $\mathscr{Q}_{\mathrm{MF}}$, and their respective associated lower transition rate operators $\underline{Q}_{\mathrm{RA}}$, $\underline{Q}_{\mathrm{LF}}$ and $\underline{Q}_{\mathrm{MF}}$.
Note that $\mathscr{Q}_{\mathrm{LF}} = \mathscr{Q}_{\mathrm{MF}} \eqqcolon \mathscr{Q}_{\mathrm{LM}}$ and $\underline{Q}_{\mathrm{LF}} = \underline{Q}_{\mathrm{MF}} \eqqcolon \underline{Q}_{\mathrm{LM}}$.
For $n_2 = 2$ there is no imprecision---that is $\underline{Q}_{\mathrm{RA}} = \tilde{Q}_{\mathrm{RA}}$ and $\underline{Q}_{\mathrm{LM}} = \tilde{Q}_{\mathrm{LM}}$---because in this case, as discussed in Section~\ref{ssec:ApprMCMod:TransitionRates}, the bounds on $\mu^{+}_{1}(i, j, e, t)$ are degenerate.

	\subsection{A Policy-Independent Imprecise MC Model}
\label{ssec:iMCMod:PolicyIndependent}
Next to elegantly handling the indeterminacy of $\mu^{+}_{1}(i, j, e, t)$, we also determine (not necessarily tight) best and worst case blocking probabilities for all possible policies.
To this end, we consider the set $\mathscr{Q}_{\mathrm{PI}}$ of transition rate matrices that are compatible with all the relevant (bounds on the) rates specified in Section~\ref{ssec:ApprMCMod:TransitionRates}, with the exception that for $R > 0$ we now only require that
\begin{equation}
\label{eqn:LambdaImprecise}
	\lambda^{-}_{\mathrm{AP}}(i, j, e, t)
	\in \begin{cases}
		\{ \lambda_{1} \} &\text{if}~ i = n_2 (m_2 - j - e), \\
		[0, \lambda_{1}] &\text{otherwise}
	\end{cases}
\end{equation}
and that $\lambda^{-}_{\mathrm{AP}}(i, j, e, t) + \lambda^{=}_{\mathrm{AP}}(i, j, e, t) = \lambda_{1}$.
In this way, we cover all policies that assign type 1 (type 2) requests if there is at least one empty (super)channel, including time-dependent policies or policies that depend on the order of the allocated flows!
The lower transition rate operator that corresponds to $\mathscr{Q}_{\mathrm{PI}}$---again using \eqref{eq:Qlower}---is denoted by $\underline{Q}_{\mathrm{PI}}$

	\subsection{Ergodicity and Limit Lower and Upper Probabilities}
\label{ssec:iMC:Ergodicity}
Each of the aforementioned lower transition operators can be easily evaluated; see Appendix~\ref{app:LTROAndLumping} for details on how to do this.
We here only mention that evaluating $\underline{Q}_{\mathrm{RA}} f$, $\underline{Q}_{\mathrm{LM}} f$ or $\underline{Q}_{\mathrm{PI}} f$ is linear in the number of states.
Indeed, for any $Q$ in $\mathscr{Q}_{\mathrm{RA}}$, $\mathscr{Q}_{\mathrm{LM}}$ or $\mathscr{Q}_{\mathrm{PI}}$ and every state $x\in\mathscr{X}_{\mathrm{red}}$, it is evident from the transition diagram of Fig.~\ref{fig:ApproximateStateSpace} that $[Q f](x)$ is a linear function with at most six non-zero coefficients, which we then need to minimise with respect to the simple inclusions in \eqref{eq:muimprecise} and, for $\underline{Q}_{PI} f$, also \eqref{eqn:LambdaImprecise}.

In Appendix~\ref{app:LTROs}, we show that $\mathscr{Q}_{\mathrm{RA}}$, $\mathscr{Q}_{\mathrm{LM}}$ and $\mathscr{Q}_{\mathrm{PI}}$ have separately specified rows and that the corresponding lower transition rate operators $\underline{Q}_{\mathrm{RA}}$, $\underline{Q}_{\mathrm{LM}}$ and $\underline{Q}_{\mathrm{PI}}$ are ergodic.
Therefore, for every $A\subseteq\mathscr{X}_{\mathrm{red}}$, there are unique lower and upper limit probabilities $\underline{\pi}_A$ and $\overline{\pi}_A$, which can be computed using the method introduced in Section~\ref{subsec:impreciseMCs}.
The resulting bounds $\underline{\pi}_A$ and $\overline{\pi}_A$ do not require us to estimate the unknown transition rates, nor do they require us to specify a spectrum allocation policy in the case of $\underline{Q}_{\mathrm{PI}}$.
For all of the above imprecise MCs, the lower/upper bounds on the blocking probability BP1 (respectively BP2) experienced by type 1 (respectively type 2) requests corresponds to choosing $A_1=\{(i,j,e)\in\mathscr{X}_{\mathrm{red}} \colon R=0\}$ (respectively $A_2=\{(i,j,e)\in\mathscr{X}_{\mathrm{red}} \colon e=0\}$).

	\section{Numerical Assessment}
\label{sec:Results}
We consider the spectrum assignment problem over a link with up to $F=480$ slices.
Type 1 channels are formed by $n_1=3$ contiguous slices (including guardbands), whereas type 2 superchannels consist of $n_1 n_2 = 3 n_2$ slices, where $n_2$ varies according to the considered scenario.

	\subsection{Scalability of the Detailed Versus the Reduced State Space}
\label{ssec:MCMod:NumericalAssesment}
We start by evaluating the number of states of the detailed state space~$\mathscr{X}_{\text{det}}$ as a function of the number of type 1 slots~$m_1$.
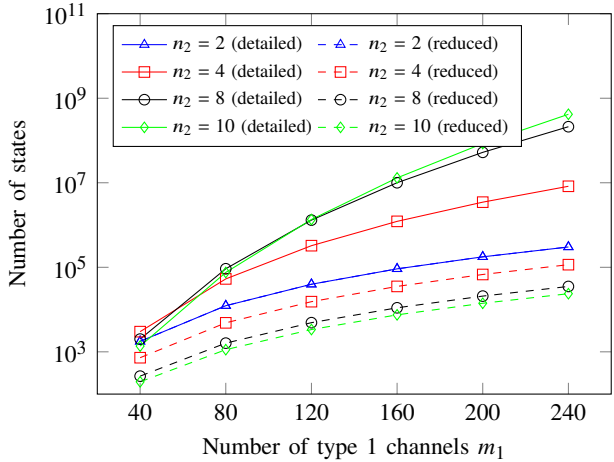
\begin{figure}[ht]
\centering
	\begin{tikzpicture}
		\pgfplotstableread{statespacecomparison.csv}{\datass}
		\begin{semilogyaxis}[
			width=.95\columnwidth,
			height=.75\columnwidth,
            label style={font=\small},
			xlabel={Number of type 1 channels $m_1$},
			ylabel={Number of states},
		    ymin=1e2, ymax=1e11,
            ticklabel style = {font=\small},
			legend entries={
				$n_2 = 2$ (detailed),
				$n_2 = 2$ (reduced),
				$n_2 = 4$ (detailed),
				$n_2 = 4$ (reduced),
				$n_2 = 8$ (detailed),
				$n_2 = 8$ (reduced),
				$n_2 = 10$ (detailed),
				$n_2 = 10$ (reduced),
			},
			legend columns=2,
			legend cell align=left,
			legend pos=north west,
			legend style={
				font=\scriptsize,
			},
			mark options={solid},
			cycle multiindex list={
				blue, blue, red, red, black, black, green, green \nextlist
				solid, dashed, solid, dashed, solid, dashed, solid, dashed\nextlist
				mark=triangle, mark=triangle, mark=square, mark=square, mark=o, mark=o, mark=diamond, mark=diamond\nextlist
			},
			xtick={40,80,120,160,200,240},
			xticklabels={40,80,120,160,200,240},
		]

			\addplot+ table [x=m1, y=2det]{\datass};
			\addplot+ table [x=m1, y=2red]{\datass};

			\addplot+ table [x=m1, y=4det]{\datass};
			\addplot+ table [x=m1, y=4red]{\datass};

			\addplot+ table [x=m1, y=8det]{\datass};
			\addplot+ table [x=m1, y=8red]{\datass};

			\addplot+ table [x=m1, y=10det]{\datass};
			\addplot+ table [x=m1, y=10red]{\datass};

		\end{semilogyaxis}
	\end{tikzpicture}
	\caption{The number of states in the detailed state space $\mathscr{X}_{\mathrm{det}}$ and the reduced state space $\mathscr{X}_{\mathrm{red}}$ as a function of $m_1$ and $n_2$.}
	\label{fig:StateSpaceComparison}
\end{figure}
As reported in Fig.~\ref{fig:StateSpaceComparison}, the number of states in the detailed state space rapidly grows with increasing $m_1$, whereas the number of states in the reduced state space~$\mathscr{X}_{\mathrm{red}}$ is at least an order of magnitude smaller if $n_2 > 2$.
Moreover, the number of states in the reduced state space grows at a lower rate than the number of states in the detailed state space for $n_2 > 2$.
Note also that, as already mentioned before, in case $n_{2} = 2$, the number of states in the detailed and reduced state space is equal.

	\subsection{Numerically Determining Precise Limit Probabilities}
\label{sssec:Numerical assesment:Numerically determining precise limit probabilties}
It is well-known that there are multiple methods for computing limit probabilities of precise MCs.
We here evaluate the performance of three such methods for the exact and approximate models described in Section~\ref{sec:MCModels}.

\paragraph{Precise version of Alg.~\ref{alg:Standard}}
The first method we consider is the precise version of Alg.~\ref{alg:Standard}, which is obtained by replacing the lower transition rate operator $\underline{Q}$ with a transition rate matrix $Q$.
This method iteratively determines the limit probability~$\pi_{A}$ of an event $A$ up to some relative tolerance $\phi$.
Note that Alg.~\ref{alg:Standard} is similar to the well-known \emph{power method} \cite{2009Stewart}, which determines the whole stationary distribution.
We opt to use the precise version of Alg.~\ref{alg:Standard} instead of the power method because this allows for easier comparison in Section~\ref{ssec:Numerical assessment:discussion on bounds}.
We implemented Alg.~\ref{alg:Standard} in Python using NumPy and SciPy functions, which are optimised for sparse matrices and dense vectors.
For the parameters, we use $\delta = 0.9 \cdot 2 / \Vert Q_{\mathrm{AP}} \Vert$, $\phi = \num{e-3}$ and $N = \num{e6}$.

\paragraph{Generalised Minimal RESidual (GMRES)}
When the limit distribution $\pi$ of an irreducible MC is interpreted as a row vector whose components are the limit probabilities $\pi_{x}$ of the states, then it is well-known---see for instance \cite{1998Norris,2009Stewart}---that $\pi$ is the unique probability mass function on $\mathscr{X}$ that satisfies the equilibrium condition $\pi Q = 0$.
As explained in \cite[Section~10.2.2]{2009Stewart}, one can determine this limit distribution by numerically solving the linear system of equations $\pi R = b$, where the matrix $R$ is obtained from $Q$ by changing the elements of the last column of $Q$ to $1$, and $b$ is a row vector where every component is zero except for the last one, which is equal to $1$.
We solve this system of equations using the LGMRES algorithm implemented in SciPy, which is a slightly adapted version of the GMRES algorithm for the sake of faster convergence \cite{2005Baker}, and the SciPy ILU function as preconditioner.
For all parameters, we simply use the standard SciPy values.

\paragraph{Gillespie}
The third method we consider provides an estimate of the blocking probabilities.
Indeed, an estimate for the limit probability $\pi_{A}$ can be obtained from a sample path $\omega(t) \colon [0, T] \to \mathscr{X}$ of the irreducible MC over a time period $T \in \mathbb{R}_{> 0}$, as the ergodic theorem for MCs guarantees that
$
	\frac{1}{T} \int_{0}^{T} \mathbb{I}_{A}(\omega(t)) \mathrm{d} t
$
converges to $\pi_{A}$ for $T \to + \infty$.
Note that the convergence of the estimate is only almost surely and that the quality of the estimate is contingent on the quality of the used random number generator.
We here generate a sample path using the Gillespie algorithm, and asses the accuracy of the approximation by means of the batch mean estimation method \cite{1990Pawlikowski}.
By the PASTA property, we only have to observe the system at the arrival epochs of a Poisson process instead of keeping track of all inter-transition times.
Therefore, we can simulate the system by generating a sample path of the embedded  (discrete-time) Markov chain (sometimes also called the jump chain).
If a transition in this sample path corresponds to the arrival of a type 1 or type 2 flow, we observe if one of the two request types would be blocked at that instant.
In our C implementation of the combination of the Gillespie algorithm and the batch mean estimation method, we use batches of \num{4e7} arrivals and \cite[Rule~1]{1990Pawlikowski} as a rule to determine the end of the burn in period.
After that, we iteratively simulate at least \num{5} and at most \num{50} batches.
If the number of batches is in between the minimum and maximum number of batches, we pre-emptively stop the simulation if the relative error---taken to be the width of the 95\%-confidence interval divided by the mean, as proposed in \cite[Eq.~(12)]{1990Pawlikowski}---is smaller than the tolerance $\phi = \num{e-3}$.

	\subsection{Determining Blocking Probabilities}
We now use each of the aforementioned computational methods---the three methods for precise MCs described in Section~\ref{sssec:Numerical assesment:Numerically determining precise limit probabilties}, and Alg.~\ref{alg:Standard} for imprecise MCs---to obtain (lower and upper bounds for) the blocking probabilities for a set of scenarios, which will allow us to make a fair assessment of the performance of each method.
The way we go about doing this is the following.
We first select four distinct systems by specifying $m_{1}$ and setting $n_{2} = 4$.
Each combination of $m_{1}$ and $n_{2}$ results in a distinct detailed state space $\mathscr{X}_{\mathrm{det}}$ and reduced state space $\mathscr{X}_{\mathrm{red}}$, the size of which is reported in Tab.~\ref{tab:Simulation scenarios}.
For every system defined this way, we set $\mu_{1} = \mu_{2} = 1$ and $\lambda_{1} = \rho \mu_{1} = \rho \mu_{2} = \lambda_{2}$, where $\rho \in \mathbb{R}_{> 0}$ is called the \emph{traffic load}.

The performance of the methods we want to compare varies with the traffic load $\rho$, which is why for every system we consider a low, medium and high traffic load, as listed in Tab.~\ref{tab:Simulation scenarios}.
Note that, as the computational complexity of the iterative approximation methods in general increases with the number of states, using the exact chains---those with transition rate matrices $Q_{\mathrm{RA}}$, $Q_{\mathrm{LF}}$ or $Q_{\mathrm{MF}}$---to determine the blocking probabilities is only feasible for relatively low values of $m_1$ (and/or $n_2$).

\begin{table}[ht]
	\renewcommand{\arraystretch}{1.3}  
	\centering
	\caption{%
		Overview of the Parameters Used in the Numerical Experiments.
	}
	\label{tab:Simulation scenarios}
	\begin{tabular}{S[table-format=3.]S[table-format=1.]S[table-format=7.]S[table-format=5.]S[table-format=1.]S[table-format=2.]S[table-format=3.]}
		\toprule
			{$m_1$} & {$n_{2}$} & {$\vert \mathscr{X}_{\mathrm{det}} \vert$} & {$\vert \mathscr{X}_{\mathrm{red}} \vert$} & {$\rho_{\mathrm{low}}$} & {$\rho_{\mathrm{med}}$} & {$\rho_{\mathrm{hi}}$} \\
		\midrule
			40 & 4 & 3003 & 726 & 2 & 10 & 50 \\
			80 & 4 &  53130 & 4851 & 8 & 28 & 100  \\
			120 & 4 & 324632 & 15376 & 16 & 50 & 150\\
			160	& 4 & 1221759 & 35301 & 32 & 80 & 200 \\
		\bottomrule
	\end{tabular}
\end{table}

\subsubsection{Exact Blocking Probabilities}
\label{ssec:Numerical assessment:Exact blocking probabilities}
For each of the three  methods discussed in Section~\ref{sssec:Numerical assesment:Numerically determining precise limit probabilties} and every system in Tab.~\ref{tab:Simulation scenarios}, we measure the execution time of the computations that are required to determine all blocking probabilities.
We go about this by fixing a numerical method, a value for $m_{2}$ and a traffic load $\rho$, and timing how long it takes to determine the blocking probabilities for all three allocations policies (corresponding to our three distinct MC models).
In order to end up with a reliable value for these timings, we determine the average of the execution time over five consequent runs.
In order to facilitate comparing the methods and models later on, for every combination of numerical method and values for $m_{1}$ and $\rho$ we actually report the average of the execution time divided by 3 (i.e., the number of distinct models).
These execution times are reported in Tab.~\ref{tab:ExactComputations}, where we also report the mean execution time over all three traffic loads.
If the (mean) execution time is longer than $120$ seconds, then we say that the computations have timed out and denote this with a slash (i.e., ``$/$'') in Tab.~\ref{tab:ExactComputations}.
\begin{table*}[ht]
	\centering
		\caption{%
			Average Execution Time (in seconds) for Determining the Exact Blocking Probabilities.}
		\label{tab:ExactComputations}
		\begin{tabular}{S[table-format=3]S[table-format=2.2]S[table-format=2.2]S[table-format=2.2]S[table-format=2.2]S[table-format=2.2]S[table-format=2.2]S[table-format=2.2]S[table-format=2.2]S[table-format=2.2]S[table-format=2.2]S[table-format=2.2]S[table-format=2.2]}
			\toprule
        	& \multicolumn{4}{c}{Alg.~\ref{alg:Standard}} & \multicolumn{4}{c}{GMRES} & \multicolumn{4}{c}{Gillespie} \\
			{$m_1$} & {$\rho_{\text{low}}$} & {$\rho_{\text{med}}$} & {$\rho_{\text{hi}}$} & {mean} & {$\rho_{\text{low}}$} & {$\rho_{\text{med}}$} & {$\rho_{\text{hi}}$} & {mean} &  {$\rho_{\text{low}}$} & {$\rho_{\text{med}}$} & {$\rho_{\text{hi}}$} & {mean} \\
			\cmidrule(r){1-1}\cmidrule(lr){2-5}\cmidrule(lr){6-9}\cmidrule(l){10-13}
			40 & 0.03 & 0.04 & 0.21 & 0.10 & 0.16 & 0.16 & 0.13 & 0.15 & / & 66.11 & 14.68 & 71.37 \\
			80 & 1.62 & 3.23 & 10.53 & 5.12 & 36.97 & 35.42 & 25.06 & 32.48 & / & 81.27 & 13.38 & 73.82 \\
			120 & 15.16 & 43.92 & 93.70 & 50.93 & / & / & / & /  & / & 85.67 & 16.10 & 78.43 \\
			160 & 100.18 & / & / & / & / & / & / & / & / & 86.00 & 14.79 & 76.35 \\
			\bottomrule
		\end{tabular}
\end{table*}
The Python code we used to obtain these numbers is available on-line\footnote{%
	\url{https://github.com/alexander-e/iCTMC-flexi-grid-allocation-policies
}
} and was run on a workstation with an Intel i5-7600 CPU @ 3.50GHz, running Ubuntu 16.04.

We observe that for Alg.~\ref{alg:Standard}, the execution time increases more than linearly with the number of states and times out for the largest system, which is in line with our expectations.
We also observe the same for the GMRES method, which actually already times out for the second largest system.
From this we conclude that using Alg.~\ref{alg:Standard} or the GMRES method to determine the blocking probabilities is only feasible for small systems.
In fact, compared to the Gillespie method, these two methods are considerably faster for small systems.
On the other hand, the Gillespie method has an execution time that does not change significantly with the model size, making it more suited for larger systems.
We also observe that the execution time for Alg.~\ref{alg:Standard} increases with increasing traffic load, while it decreases for GMRES and Gillespie.
Even more, the Gillespie method is not suited for low traffic loads since it times out for these loads regardless of the model size.
A more accurate comparison is hard to make because the execution times are dependent on implementation choices.

\subsubsection{Approximate Blocking Probabilities}
\label{ssec:Numerical assessment:Approxaimate blocking probabilities}
We now run the same computations for the precise MCs characterised by $\tilde{Q}_{\mathrm{RA}}$ and $\tilde{Q}_{\mathrm{LM}}$.
Recall that for every combination of $m_{1}$ and $\rho$, we report the average of the execution time over five consecutive runs divided by the number of MC models, in this case 2.
We report our findings in Tab.~\ref{tab:ApproximateAndBoundsComputations}.
Note that we do not consider the Gillespie method here because its implementation is slightly less straightforward in this case and because the two alternative methods for precise MC models are already sufficiently fast.
\begin{table*}[ht]
	\centering
		\caption{%
			Average Execution Time (in seconds) for Determining Approximations of and Bounds on the Blocking Probabilities.}
		\label{tab:ApproximateAndBoundsComputations}
		\begin{tabular}{S[table-format=3]S[table-format=2.2]S[table-format=2.2]S[table-format=2.2]S[table-format=2.2]S[table-format=2.2]S[table-format=2.2]S[table-format=2.2]S[table-format=2.2]S[table-format=2.2]S[table-format=2.2]S[table-format=2.2]S[table-format=2.2]S[table-format=2.2]S[table-format=2.2]S[table-format=2.2]S[table-format=2.2]}
			\toprule
			& \multicolumn{8}{c}{Approximate blocking probabilities} & \multicolumn{8}{c}{Bounds on the blocking probabilities} \\
			\cmidrule(r){2-9}\cmidrule(l){10-17}
        	& \multicolumn{4}{c}{Alg.~\ref{alg:Standard}} & \multicolumn{4}{c}{GMRES} & \multicolumn{4}{c}{Policy-dependent} & \multicolumn{4}{c}{Policy-independent} \\
			{$m_1$} & {$\rho_{\text{low}}$} & {$\rho_{\text{med}}$} & {$\rho_{\text{hi}}$} & {mean} & {$\rho_{\text{low}}$} & {$\rho_{\text{med}}$} & {$\rho_{\text{hi}}$} & {mean} & {$\rho_{\text{low}}$} & {$\rho_{\text{med}}$} & {$\rho_{\text{hi}}$} & {mean} & {$\rho_{\text{low}}$} & {$\rho_{\text{med}}$} & {$\rho_{\text{hi}}$} & {mean} \\
			\cmidrule(r){1-1}\cmidrule(lr){2-5}\cmidrule(lr){6-9}\cmidrule(lr){10-13}\cmidrule(l){14-17}
			40  & 0.02 & 0.02 & 0.11 & 0.05 & 0.01 & 0.01 & 0.01 & 0.01 & 0.16 & 0.23 & 0.92 & 0.44 & 0.22 & 0.25 & 0.97 & 0.48 \\
			80  & 0.13 & 0.21 & 0.87 & 0.40 & 0.17 & 0.16 & 0.14 & 0.16 & 1.24 & 3.67 & 6.83 & 3.91 & 2.01 & 2.49 & 9.53 & 4.68 \\
			120 & 0.60 & 1.20 & 3.97 & 1.92 & 1.04 & 0.87 & 0.75 & 0.89 & 6.95 & 52.47 & 37.20 & 32.21 & 11.6 & 16.46 & 55.16 & 27.74 \\
			160 & 2.08 & 4.82 & 11.69 & 6.20 & 3.82 & 3.52 & 2.32 & 3.22 & 42.58 & / & 111.43 & / & 39.93 & 60.73 & / & 84.07 \\
			\bottomrule
		\end{tabular}
		\vspace{1em}\\
		\hrulefill
\end{table*}
By comparing the results to those of Tab.~\ref{tab:ExactComputations}, we observe that the decrease in the number of states---see Tab.~\ref{tab:Simulation scenarios}---indeed results in a substantial decrease of the required computational time.

\subsubsection{Lower and Upper Blocking Probabilities}
\label{ssec:Numerical assessment:lower and upper blocking probabilities}
Finally, we determine lower and upper bounds on the blocking probabilities using the imprecise MC models introduced in Section~\ref{sec:iMCModels}.
These bounds are determined using Alg.~\ref{alg:Standard}, with $\delta = 1 / (2 \Vert \underline{Q}_{\mathrm{AP}} \Vert)$, $\phi = 10^{-3}$ and $N = 10^6$.
For the two policy-dependent imprecise MC models---characterised by $\underline{Q}_{\mathrm{RA}}$ and $\underline{Q}_{\mathrm{LM}}$---we report the average execution time over five consecutive runs divided by 2, whereas for the policy-independent MC model---characterised by $\underline{Q}_{\mathrm{PI}}$---we simply report the average execution time over five consecutive runs.
In Tab.~\ref{tab:ApproximateAndBoundsComputations}, we report the obtained timings.

Comparing the mean execution times of the policy-dependent and policy-independent MCs, we observe that those of the policy-independent MC are lower for large $m_{1}$.
This is rather surprising because---for reasons explained in Appendix~\ref{app:LTROAndLumping}---an iteration for the policy-independent MC requires more computations than an iteration for the policy-dependent MCs.
Comparing the average execution times per load, we see that for the low and high load these are shorter for the policy-dependent models than for the policy-independent model, as expected.
However, the average execution time for the medium traffic load and the larger systems is longer for the policy-dependent models than for the policy-independent model.
Taking a more detailed look at the raw data, we find that this is caused by the extremely slow convergence of the lower and upper bounds for the medium traffic load and the imprecise MC model characterised by $\underline{Q}_{\text{LM}}$ (for example, computing all bounds on blocking probabilities for $m_{1} = 160$, $\rho_{\mathrm{med}}$ and $\underline{Q}_{\mathrm{LM}}$ took about 18 minutes).
We currently have no intuitive explanation for this behaviour and consider resolving these convergence issues a matter for further study.

For the two smallest systems and the policy-dependent MCs, the mean execution time is approximately equal to 10 times the mean execution time for the approximate MC models in Tab.~\ref{tab:ApproximateAndBoundsComputations}, which is as expected.
Similarly, for all systems and the policy-independent MCs, the mean execution time is about 10 times the mean execution time for the approximate MC models or approximately equal to the mean execution time for the policy-dependent MCs.
This is faster than expected, because---as explained in Appendix~\ref{app:LTROAndLumping}---the policy-independent MCs require more computations per iteration than the policy-dependent MCs.

	\subsection{Accuracy of the Precise But Approximate MC Models}
\label{ssec:Numerical assessment:Accuracy of approximate}
\begin{figure*}[p]
\subfloat[RA policy \label{fig:PreciseRA}]{
	\begin{tikzpicture}
		\pgfplotstableread{m1_80_n2_4.csv}{\datasmall}
		\begin{loglogaxis}[
			width=.48\textwidth,
			height=.35\textwidth,
            label style={font=\small},
			xlabel={Offered Traffic Load},
			ylabel={Blocking Probability},
		    ymin=1e-8, ymax=2,
		    xtick={8, 28, 100},
		    xticklabels={8, 28, 100},
		    transpose legend,
			legend entries={
				BP1 (exact),
				BP1 (approximate),
				BP2 (exact),
				BP2 (approximate),
			},
			legend columns=2,
			legend cell align=left,
			legend pos=south east,
			legend style={
				font=\scriptsize,
			},
			mark options={solid},
			cycle multiindex list={
				blue, blue, red, red \nextlist
				solid, solid, dashed, dashed\nextlist
				mark=none, mark=o, mark=none, mark=square\nextlist
			}
		]
		\addplot+ table [x=rho, y=RA:BP1]{\datasmall};
		\addplot+ [only marks] table [x=rho, y=R:BP1]{\datasmall};

		\addplot+ table [x=rho, y=RA:BP2]{\datasmall};
		\addplot+ [only marks] table [x=rho, y=R:BP2]{\datasmall};
		\end{loglogaxis}
	\end{tikzpicture}
}\hfil
\subfloat[LF and MF policies \label{fig:PreciseLFMF}]{
	\begin{tikzpicture}
	\pgfplotstableread{m1_80_n2_4.csv}{\datasmall}
		\begin{loglogaxis}[
			width=.48\textwidth,
			height=.35\textwidth,
            label style={font=\small},
			xlabel={Offered Traffic Load},
			ylabel={Blocking Probability},
			ymin=1e-8, ymax=2,
			xtick={8, 28, 100},
		    xticklabels={8, 28, 100},
			transpose legend,
			legend columns=3,
			legend cell align=left,
			legend entries={
				{BP1 (LF, exact)},
				{BP1 (MF, exact)},
				BP1 (approximate),
				{BP2 (LF, exact)},
				{BP2 (MF, exact)},
				BP2 (approximate),
			},
			legend pos=south east,
			legend style={
				font=\scriptsize,
			},
			cycle multiindex list={
				blue, blue, blue, red, red, red \nextlist
				solid, dashdotted, solid, dashed, dotted, solid\nextlist
				mark=none, mark=none, mark=o, mark=none, mark=none, mark=square\nextlist
			}
		]
			\addplot+ table [x=rho, y=LF:BP1]{\datasmall};
			\addplot+ table [x=rho, y=MF:BP1]{\datasmall};
			\addplot+ [only marks] table [x=rho, y=LM:BP1]{\datasmall};

			\addplot+ table [x=rho, y=LF:BP2]{\datasmall};
			\addplot+ table [x=rho, y=MF:BP2]{\datasmall};
			\addplot+ [only marks] table [x=rho, y=LM:BP2]{\datasmall};
		\end{loglogaxis}
	\end{tikzpicture}
    }
	\caption{Blocking probabilities for a system with $m_1 = 80$ and $n_2=4$, for the Random Allocation (RA), Least Filled (LF) and Most Filled (MF) allocation policies.}
\end{figure*}
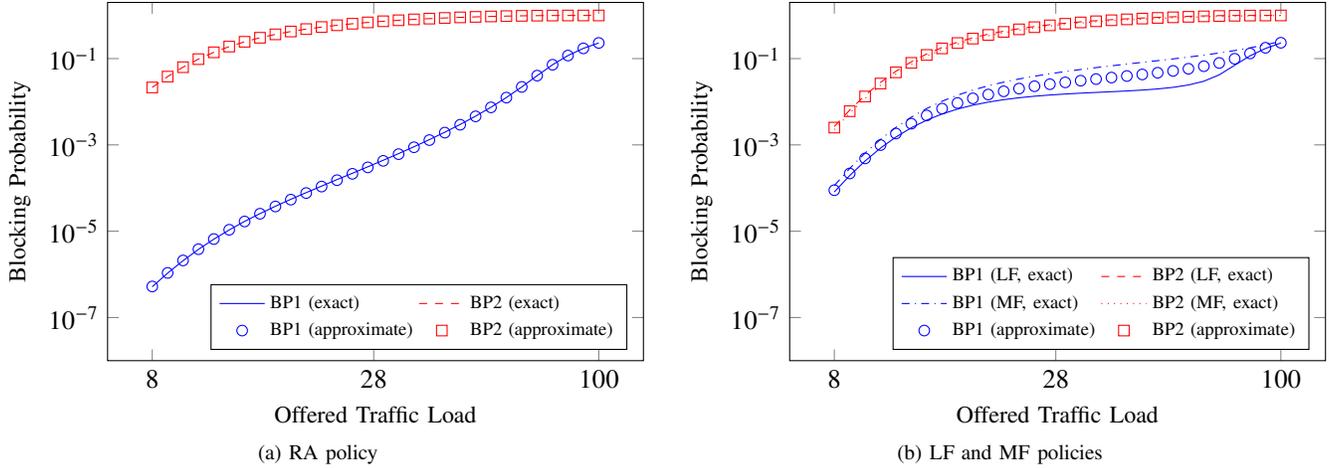

In Figs.~\ref{fig:PreciseRA} and \ref{fig:PreciseLFMF}, we depict the blocking probabilities calculated for the same system with $m_{1} = 80$ and $n_{2} = 4$ using the RA policy or the LF and MF allocation policies, respectively.
From Fig.~\ref{fig:PreciseRA}, we conclude that for the RA policy the blocking probabilities obtained using the approximate MC model are in good accordance with those obtained using the exact MC model.
This is as expected, as it was previously observed by \citet{kim2015analytical}.
However, for the LF and MF allocation policies, we observe something else.
On the one hand, these two policies have a different performance, which can be evaluated with their respective exact MCs.
On the other hand, the approximate model yields identical approximate blocking probabilities for the two policies.
From Fig.~\ref{fig:PreciseLFMF}, we conclude that the approximations of BP2 for the LF and MF policies are in good accordance with the actual values of BP2.
However, it also becomes evident that there is a range of traffic loads for which the approximation for BP1 is not very good.
We conclude that the approximate MC models are affected by approximation errors that may result in incorrect performance evaluation and system dimensioning.
Therefore, rigorous bounds are necessary to make correct design choices.

	\subsection{Discussion on Probability Bounds Provided by the Imprecise MC Models}
\label{ssec:Numerical assessment:discussion on bounds}
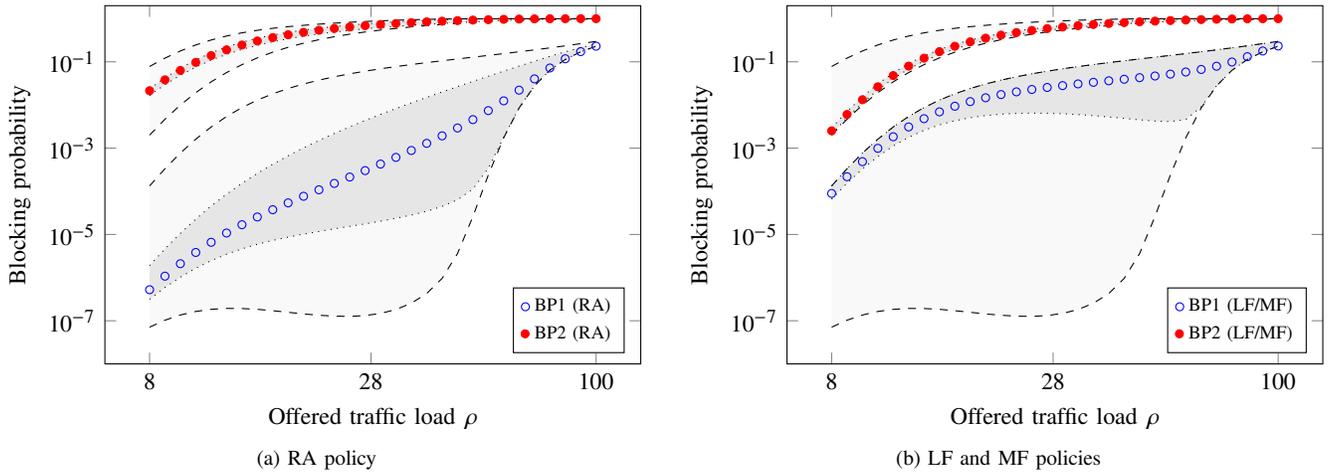
\begin{figure*}[p]
\subfloat[RA policy \label{fig:boundsra80}]{
\begin{tikzpicture}
	\pgfplotstableread{m1_80_n2_4.csv}{\data}
		\begin{loglogaxis}[
			width=.48\textwidth,
			height=.35\textwidth,
            label style={font=\small},
			xlabel={Offered traffic load $\rho$},
			ylabel={Blocking probability},
			ymin=1e-8, ymax=2,
            ticklabel style = {font=\small},
			xtick={8, 28, 100},
			xticklabels={8, 28, 100},
			mark options={solid},
			mark size={1.5pt},
			cycle multiindex list={
				black, black, blue, black, black,
				black, black, red, black, black,\nextlist
				dashed, dotted, solid, dotted, dashed,
				dashed, dotted, solid, dotted, dashed\nextlist
				mark=none, mark=none, mark=o, mark=none, mark=none,
				mark=none, mark=none, mark=*, mark=none, mark=none\nextlist
			},
			legend entries={
				,,BP1 (RA),,,
				,,BP2 (RA),,,
			},
			legend cell align=left,
			legend pos=south east,
			legend style={
				font=\scriptsize,
			},
		]
			\addplot+ [name path=A1] table [x=rho, y=I:BP1L]{\data};
			\addplot+ [name path=A2] table [x=rho, y=R:BP1L]{\data};
			\addplot+ [only marks] table [x=rho, y=R:BP1]{\data};
			\addplot+ [name path=B2] table [x=rho, y=R:BP1U]{\data};
			\addplot+ [name path=B1] table [x=rho, y=I:BP1U]{\data};

			\addplot+ [name path=C1] table [x=rho, y=I:BP2L]{\data};
			\addplot+ [name path=C2] table [x=rho, y=R:BP2L]{\data};
			\addplot+ [only marks] table [x=rho, y=R:BP2]{\data};
			\addplot+ [name path=D2] table [x=rho, y=R:BP2U]{\data};
			\addplot+ [name path=D1] table [x=rho, y=I:BP2U]{\data};

			\addplot[gray!5] fill between[of=A1 and B1];
			\addplot[gray!5] fill between[of=C1 and D1];
			\addplot[gray!20] fill between[of=A2 and B2];
			\addplot[gray!20] fill between[of=C2 and D2];
		\end{loglogaxis}
	\end{tikzpicture}
}\hfil
\subfloat[LF and MF policies \label{fig:boundslfmf80}]{
	\begin{tikzpicture}
	\pgfplotstableread{m1_80_n2_4.csv}{\data}
		\begin{loglogaxis}[
			width=.48\textwidth,
			height=.35\textwidth,
            label style={font=\small},
			xlabel={Offered traffic load $\rho$},
			ylabel={Blocking probability},
			ymin=1e-8, ymax=2,
            ticklabel style = {font=\small},
			xtick={8, 28, 100},
			xticklabels={8, 28, 100},
			mark options={solid},
			mark size={1.5pt},
			cycle multiindex list={
				black, black, blue, black, black,
				black, black, red, black, black,\nextlist
				dashed, dotted, solid, dotted, dashed,
				dashed, dotted, solid, dotted, dashed\nextlist
				mark=none, mark=none, mark=o, mark=none, mark=none,
				mark=none, mark=none, mark=*, mark=none, mark=none\nextlist
			},
			legend entries={
				,,BP1 (LF/MF),,,
				,,BP2 (LF/MF),,,
			},
			legend cell align=left,
			legend pos=south east,
			legend style={
				font=\scriptsize,
			},
		]
			\addplot+ [name path=A1] table [x=rho, y=I:BP1L]{\data};
			\addplot+ [name path=A2] table [x=rho, y=LM:BP1L]{\data};
			\addplot+ [only marks] table [x=rho, y=LM:BP1]{\data};
			\addplot+ [name path=B2] table [x=rho, y=LM:BP1U]{\data};
			\addplot+ [name path=B1] table [x=rho, y=I:BP1U]{\data};

			\addplot+ [name path=C1] table [x=rho, y=I:BP2L]{\data};
			\addplot+ [name path=C2] table [x=rho, y=LM:BP2L]{\data};
			\addplot+ [only marks] table [x=rho, y=LM:BP2]{\data};
			\addplot+ [name path=D2] table [x=rho, y=LM:BP2U]{\data};
			\addplot+ [name path=D1] table [x=rho, y=I:BP2U]{\data};

			\addplot[gray!5] fill between[of=A1 and B1];
			\addplot[gray!5] fill between[of=C1 and D1];
			\addplot[gray!20] fill between[of=A2 and B2];
			\addplot[gray!20] fill between[of=C2 and D2];
		\end{loglogaxis}
	\end{tikzpicture}
	}
	\caption{Blocking probabilities for a system with $m_1 = 80$ and $n_2 = 4$.
	The lower and upper bounds obtained with $\underline{Q}_{\mathrm{RA}}$ (respectively $\underline{Q}_{\mathrm{LM}}$) are displayed as dotted lines, those obtained with $\underline{Q}_{\mathrm{PI}}$ are displayed as dashed lines.}
	\label{fig:m1_80_n2}
\end{figure*}
\begin{figure*}[p]
\subfloat[RA policy]{\label{fig:boundsra120}
\begin{tikzpicture}
	\pgfplotstableread{m1_120_n2_4.csv}{\data}
		\begin{loglogaxis}[
			width=.48\textwidth,
			height=.35\textwidth,
            label style={font=\small},
			xlabel={Offered traffic load $\rho$},
			ylabel={Blocking probability},
			ymin=1e-11, ymax=2,
            ticklabel style = {font=\small},
			xtick={16, 50, 150},
			xticklabels={16, 50, 150},
			mark options={solid},
			mark size={1.5pt},
			cycle multiindex list={
				black, black, blue, black, black,
				black, black, red, black, black,\nextlist
				dashed, dotted, solid, dotted, dashed,
				dashed, dotted, solid, dotted, dashed\nextlist
				mark=none, mark=none, mark=o, mark=none, mark=none,
				mark=none, mark=none, mark=*, mark=none, mark=none\nextlist
			},
			legend entries={
				,,BP1 (RA),,,
				,,BP2 (RA),,,
			},
			legend cell align=left,
			legend pos=south east,
			legend style={
				font=\scriptsize,
			},
		]
			\addplot+ [name path=A1] table [x=rho, y=I:BP1L]{\data};
			\addplot+ [name path=A2] table [x=rho, y=R:BP1L]{\data};
			\addplot+ [only marks] table [x=rho, y=R:BP1]{\data};
			\addplot+ [name path=B2] table [x=rho, y=R:BP1U]{\data};
			\addplot+ [name path=B1] table [x=rho, y=I:BP1U]{\data};

			\addplot+ [name path=C1] table [x=rho, y=I:BP2L]{\data};
			\addplot+ [name path=C2] table [x=rho, y=R:BP2L]{\data};
			\addplot+ [only marks] table [x=rho, y=R:BP2]{\data};
			\addplot+ [name path=D2] table [x=rho, y=R:BP2U]{\data};
			\addplot+ [name path=D1] table [x=rho, y=I:BP2U]{\data};

			\addplot[gray!5] fill between[of=A1 and B1];
			\addplot[gray!5] fill between[of=C1 and D1];
			\addplot[gray!20] fill between[of=A2 and B2];
			\addplot[gray!20] fill between[of=C2 and D2];
		\end{loglogaxis}
	\end{tikzpicture}
}\hfil
\subfloat[LF and MF policies]{\label{fig:boundslfmf120}
	\begin{tikzpicture}
	\pgfplotstableread{m1_120_n2_4.csv}{\data}
		\begin{loglogaxis}[
			width=.48\textwidth,
			height=.35\textwidth,
            label style={font=\small},
			xlabel={Offered traffic load $\rho$},
			ylabel={Blocking probability},
			ymin=1e-11, ymax=2,
            ticklabel style = {font=\small},
			xtick={16, 50, 150},
			xticklabels={16, 50, 150},
			mark options={solid},
			mark size={1.5pt},
			cycle multiindex list={
				black, black, blue, black, black,
				black, black, red, black, black,\nextlist
				dashed, dotted, solid, dotted, dashed,
				dashed, dotted, solid, dotted, dashed\nextlist
				mark=none, mark=none, mark=o, mark=none, mark=none,
				mark=none, mark=none, mark=*, mark=none, mark=none\nextlist
			},
			legend entries={
				,,BP1 (LF/MF),,,
				,,BP2 (LF/MF),,,
			},
			legend cell align=left,
			legend pos=south east,
			legend style={
				font=\scriptsize,
			},
		]
			\addplot+ [name path=A1] table [x=rho, y=I:BP1L]{\data};
			\addplot+ [name path=A2] table [x=rho, y=LM:BP1L]{\data};
			\addplot+ [only marks] table [x=rho, y=LM:BP1]{\data};
			\addplot+ [name path=B2] table [x=rho, y=LM:BP1U]{\data};
			\addplot+ [name path=B1] table [x=rho, y=I:BP1U]{\data};

			\addplot+ [name path=C1] table [x=rho, y=I:BP2L]{\data};
			\addplot+ [name path=C2] table [x=rho, y=LM:BP2L]{\data};
			\addplot+ [only marks] table [x=rho, y=LM:BP2]{\data};
			\addplot+ [name path=D2] table [x=rho, y=LM:BP2U]{\data};
			\addplot+ [name path=D1] table [x=rho, y=I:BP2U]{\data};

			\addplot[gray!5] fill between[of=A1 and B1];
			\addplot[gray!5] fill between[of=C1 and D1];
			\addplot[gray!20] fill between[of=A2 and B2];
			\addplot[gray!20] fill between[of=C2 and D2];
		\end{loglogaxis}
	\end{tikzpicture}
    }
	\caption{Blocking probabilities for a system with $m_1 = 120$ and $n_2 = 4$.
	The lower and upper bounds obtained with $\underline{Q}_{\mathrm{RA}}$ (respectively $\underline{Q}_{\mathrm{LM}}$) are displayed as dotted lines, those obtained with $\underline{Q}_{\mathrm{PI}}$ are displayed as dashed lines.}
	\label{fig:m1_120_n2}
\end{figure*}
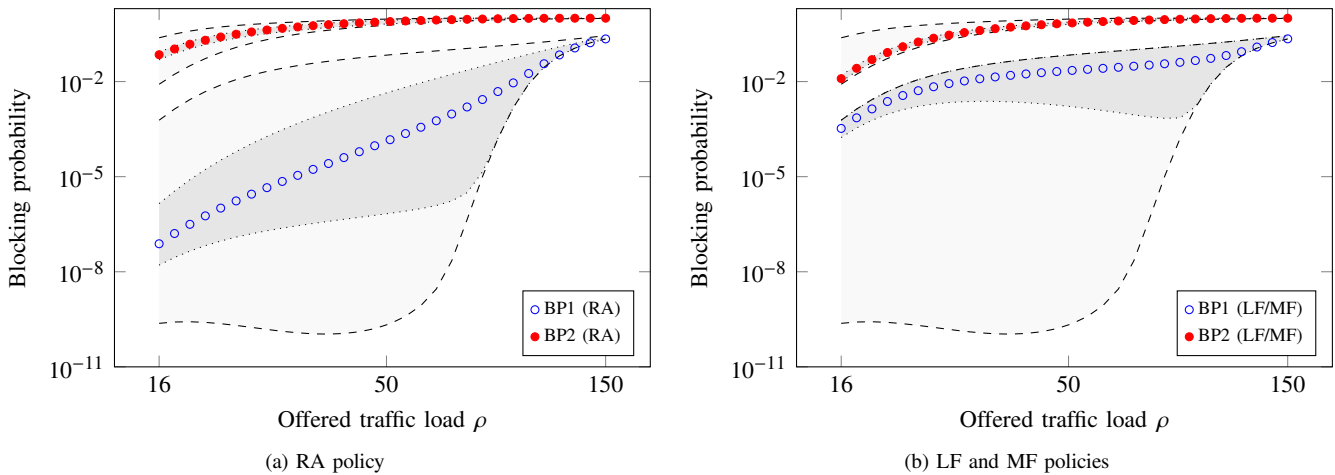

In Fig.~\ref{fig:m1_80_n2}, for each connection type---that is, for type 1 and type 2 flows---and for $m_1=80$, we show the blocking probabilities of the RA policy and of the MF/LF policies, all obtained with the approximate MC.
We show similar results for $m_1=120$ in Fig.~\ref{fig:m1_120_n2}.
In addition, the same figures show lower and upper bounds obtained with $\underline{Q}_{\mathrm{RA}}$ or $\underline{Q}_{\mathrm{LM}}$---displayed as dotted lines---as well as policy-independent bounds obtained with $\underline{Q}_{\mathrm{PI}}$---displayed as dashed lines.
Note that the graphs in Figs.~\ref{fig:m1_80_n2} and \ref{fig:m1_120_n2} all have a double logarithmic scale.
Consequently, our discussion of the graphs---and our use of terms such as tight, wide, close to or in the middle---should be interpreted in a logarithmic sense.

Along with the approximate blocking probability, the calculated bounds make it possible to evaluate the performance of the system without solving the exact model.
For type 2 flows, the bounds calculated with the policy-dependent model are very tight and show that evaluating the performance with the reduced-state model yields accurate results for all the considered parameters and traffic loads.

Different considerations hold for type 1 flows.
We know, from our comparison with the exact solution in Fig.~\ref{fig:PreciseRA}, that for $m_{1} = 80$ the reduced-state model for the RA policy is accurate.
Nevertheless, the calculated policy-dependent lower and upper bounds are relatively wide, especially in the intermediate loads.
In contrast, for the LF/MF model, we know already that the reduced-state model approximates multiple policies with different performance, and we therefore expect that the bounds cannot be very tight.
However, the bounds calculated with $\underline{Q}_{\text{LM}}$ are narrower than those calculated with $\underline{Q}_{\text{RA}}$.

By observing the bounds of the policy-independent model in Figs.~\ref{fig:boundslfmf80} and \ref{fig:boundslfmf120}, we can see that the MF and LF policies yield a blocking probability very close to the policy-independent lower bound for type 2 flows and very close to the policy-independent upper bound for type 1 flows.
This demonstrates that these policies favour type 2 flows.
On the other hand, Figures~\ref{fig:boundsra80} and \ref{fig:boundsra120} show that the RA policy is situated more or less in the middle of the two bounds, for both connection types.
This allows us to infer that the RA policy shows no preference for one type of flow over the other.

	\section{Conclusion}
\label{sec:conclusions}
We have studied the problem of spectrum assignment in a two-service flexi-grid optical link.
First, we provided exact Markov chain models for the cases of random, most-filled, and least-filled allocation policies.
We observed that the number of states in these models scales exponentially with the dimensions of the system under study.
An unfortunate consequence of this exponential scaling is that, for two out of the three approximation methods we consider, the computations required to determine the blocking probabilities---the performance measures we are interested in---become intractable for large systems.
Furthermore, our experiments indicate that the third considered approximation method is not suited to systems with low loads.
Therefore, we considered several alternative models that allow to tractably determine the blocking probabilities.
We evaluated an approximate, reduced-state Markov chain model that is available in the literature for a random allocation policy, and extended it to the case of the most-filled and least-filled policies.
For a small-scale case, we observed that the blocking probabilities obtained using these approximate models were in good accordance with the exact results for the random allocation policy, but that they were not that accurate for the least-filled and most-filled policies.
Furthermore, it is difficult to obtain an evaluation of the precision of the approximated blocking probabilities for large scale problems.
For this reason, finally, we introduced four imprecise Markov models: one model for each of the three considered allocation policies and one policy-independent model.
These models scale to large systems and provide guaranteed lower and upper bounds on the blocking probabilities instead of a single precise estimate without any measure of accuracy.
Hence, the policy-dependent imprecise Markov models can be used to evaluate the precision of their approximate counterparts for large systems.
Furthermore, the policy-independent model, which does not make assumptions about the allocation policy, provides guaranteed bounds on the achievable blocking probabilities.
As such, for a given system, this policy-independent model can be used to assess the performance of a specific policy with respect to the set of all possible policies.

\appendices

\section*{Acknowledgements}
Much of Alexander Erreygers's research was conducted while he was a member of the SMACS Research Group of the Department of Telecommunications and Information Processing at Ghent University, and he would therefore like to acknowledge their support.
Jasper~De~Bock's contribution was partially supported by the H2020-MSCA-ITN-2016 UTOPIAE, grant agreement 722734.

\bibliographystyle{IEEEtranN}
\bibliography{biblio}

	\section{Ergodicity of Transition Rate Matrices}
\label{app:PreciseErgodicity}
Let $Q$ be a transition rate matrix with state space $\mathscr{X}$.
One well-known necessary and sufficient condition for the ergodicity of $Q$ is based on the reflexive and transitive binary accessibility relation $\cdot \acces \cdot$.
For any $x, y \in \mathscr{X}$, we say that $x$ is accessible from $y$, denoted by $y \acces x$, if (i) $x = y$, or (ii) $x \neq y$ and there is a sequence $y = x_{0}, x_{1}, \dots, x_{n} = x$ in $\mathscr{X}$ such that $Q(x_{i-1}, x_{i}) > 0$ for all $i \in \{ 1, \dots, n\}$.
Then $Q$ is irreducible if and only if it is top class
\[
	\mathscr{R}_{Q}
	\coloneqq \left\{ x \in \mathscr{X} \colon (\forall y \in \mathscr{X})
	~ y \acces x \right\}
\]
is equal to the state space $\mathscr{X}$, see \cite[Chapter~3]{1998Norris}.

	\subsection{Irreducibility of the Exact Transition Rate Matrices}
\label{app:PreciseErgodicity:ErgodicityExact}
In Section~\ref{ssec:MC Models:Exact model}, we defined a transition rate matrix $Q_{\mathrm{AP}}$ with state space $\mathscr{X}_{\mathrm{det}}$ for three assignment policies AP: RA, LF and MF.
We posit that for all three assignment policies, the associated transition rate matrix $Q_{\mathrm{AP}}$ is irreducible.
This is true if for every two states $x,y \in \mathscr{X}_{\mathrm{det}}$, $y \acces x$.

Instead of proving this, we will prove that for any state $x^{\star} = (i^{\star}_{0}, \dots, i^{\star}_{n_{2}}) \in \mathscr{X}_{\mathrm{det}}$, any state of the form $x_{k}^{+} \coloneqq (i^{\star}_0, \dots, i^{\star}_k + 1, \dots i^{\star}_{n_2}) \in \mathscr{X}_{\mathrm{det}}$ can be accessed, and similarly for any $x_{k}^{-} \coloneqq (i^{\star}_0, \dots, i^{\star}_k - 1, \dots i^{\star}_{n_2}) \in \mathscr{X}_{\mathrm{det}}$.

If this is the case, then using the transitivity of the accesiblity relation we also have that from any state $y \coloneqq (j_{0}, \dots, j_{n_{2}}) \in \mathscr{X}_{\mathrm{det}}$ we can acces any other state $x \coloneqq (i_{0}, \dots, i_{n_{2}}) \in \mathscr{X}_{\mathrm{det}}$ as follows.
Let $K_{-}(y, x) \coloneqq \{ k \in [0, n_2] \colon j_{k} > i_{k} \}$ and $K_{+}(y, x) \coloneqq \{ k \in [0, n_2] \colon j_{k} < i_{k} \}$, then first for all $k \in K_{-}(y, x)$ we decrement the $k$-th component of the state by $j_k - i_k$, and second for all $k \in K_{+}(y, x)$ we increment the $k$-th component of the state by $i_k - j_k$.
This detour ensures that any tuple along the way is a state, as any such tuple is dominated component wise by either $x$ or $y$, and any tuple of non-negative natural numbers that is dominated component wise by a state is also a state.
Hence, we have that $y \acces x$; and as $x$ and $y$ were arbitrary distinct states, we can conclude that the transition rate operator $Q_{\mathrm{AP}}$ is indeed irreducible.

As the rate $\lambda_{\mathrm{AP}}$ in the transition diagram of Fig.~\ref{fig:ExactStateSpace} is policy dependent, we need to treat every policy separately.
We here only formally prove the irreducibility of $Q_{\mathrm{RA}}$.
The irreducibility of $Q_{\mathrm{LF}}$ and $Q_{\mathrm{MF}}$ can be proven in a similar fashion.
The part of the proof where we argue that $x^{\star} \acces x_{0}^{+}$ and $x^{\star} \acces x_{k}^{-}$ for $k \in [0, n_{2}]$ remains the same, it is only the part of the proof where we show that $x^{\star} \acces x_{k}^{+}$ for $k \in [1, n_{2}]$ that needs to be modified.

	\subsection{Irreducibility of \texorpdfstring{$Q_{\mathrm{RA}}$}{QRA}}
\label{app:PreciseErgodicity:RA}

Fix some state $x^{\star} \in \mathscr{X}_{\mathrm{det}}$ and some $k \in [0, n_{2}]$ such that $x^{+}_{k} \in  \mathscr{X}_{\mathrm{det}}$ and/or $x^{-}_{k} \in \mathscr{X}_{\mathrm{det}}$.
We now prove that $x^{\star} \acces x^{+}_{k}$ and/or $x^{\star} \acces x^{-}_{k}$.
Observe that for $x^{\star}$ the number of superchannels not occupied by a type 2 flow is $I^{\star} \coloneqq \sum_{k = 0}^{n_2} i^{\star}_k \leq m_2$.

First, we consider the case $k  = 0$.
From the feasibility conditions it follows that $x^{+}_{0} \in \mathscr{X}_{\mathrm{det}}$ if and only if $I^{\star} < m_{2}$.
If this is the case, then
\[
	Q_{\mathrm{RA}}(x^{\star}, x^{+}_{0}) = (m_2 - I^{\star}) \mu_2 > 0,
\]
where the strict positivity follows from $I^{\star} < m_2$ and $\mu_{2} > 0$.
We infer from this that in this case indeed $x^{\star} \acces x^{+}_{0}$.
Also, $x^{-}_{0} \in \mathscr{X}_{\mathrm{det}}$ if and only if $i^{\star}_0 > 0$.
In this case
\[
	Q_{\mathrm{RA}}(x^{\star}, x^{-}_{0}) = \lambda_2 > 0,
\]
such that indeed $x^{\star} \acces x^{-}_{0}$.

Next, we consider the case $k > 0$.
The state $x^{+}_{k}$ is then feasible if and only if $I^{\star} < m_{2}$.
In this case, we can also immediately verify that $x^{+}_{\ell}$ is a proper state for all $\ell \in \{ 0, \dots, k-1 \}$.
Recall from before that as $I^{\star} < m_{2}$,
\[
	Q_{\mathrm{RA}}(x^{\star}, x^{+}_{0}) = (m_{2} - I^{\star})\mu_{2}  > 0.
\]
Next, let $R^{\star} \coloneqq \sum_{k=0}^{n_2} i_{k}^{\star}(n_2 - k)$ and observe that for all $\ell \in \{ 0, \dots, k-1 \}$,
\[
	Q_{\mathrm{RA}}(x^{+}_{\ell}, x^{+}_{\ell+1}) = \lambda_{1} \frac{(i^{\star}_{\ell} + 1) (n_{2} - \ell)}{R^{\star} + n_{2} - \ell} > 0,
\]
where the inequality follows from $\lambda_{1} > 0$, $i^{\star}_{\ell} \geq 0$, $\ell < k \leq n_{2}$ and $R^{\star} \geq 0$.
We infer from this that in this case $x^{\star} \acces x^{+}_{k}$.

Also, $x^{-}_{k}$ is feasible if and only if $i^{\star}_k > 0$, and we now assume that this is indeed the case.
For the sake of brevity, for all $\ell \in \{ 0, \dots, k-1 \}$ we denote by $x'_{\ell}$ the $(n_{2}+1)$-tuple of non-negative natural numbers that is obtained from $x^{\star}$ by decrementing the $i^{\star}_{k}$ component by 1 and incrementing the $i^{\star}_{\ell}$ component by 1.
As we have assumed that $i^{\star}_{k} > 0$, all states $x'_{k-1}, \dots x'_{0}$ are feasible.
We now prove that in this case $x^{\star} \acces x'_{k-1} \acces \dots \acces x'_{0} \acces x^{-}_{k}$, which implies $x^{\star} \acces x^{-}_{k}$ by transitivity.
First, observe that
\[
	Q_{\mathrm{RA}}(x^{\star}, x'_{k-1}) = k i^{\star}_{k} \mu_{1} > 0,
\]
where the inequality follows from $k \geq 1$, $i^{\star}_{k} \geq 1$ and $\mu_{1} > 0$.
Second, observe that for all $\ell \in \{ 1, \dots, k-1 \}$,
\[
	Q_{\mathrm{RA}}(x'_{\ell}, x'_{\ell-1}) = \ell (i^{\star}_{\ell}+1) \mu_{1} \geq \mu_1 > 0,
\]
where the inequality follows from $\ell \geq 1$ and $i^{\star}_{\ell} \geq 0$.
Third, observe that
\[
	Q_{\mathrm{RA}}(x'_{0}, x^{-}_{k}) = \lambda_{2} > 0.
\]
From these three observations, we infer that indeed $x^{\star} \acces x^{-}_{k}$.

	\subsection{Irreducibility of the Approximate Transition Rate Matrices}
For each of the considered assignment policies AP, the irreducibility of the approximate transition rate matrix $\tilde{Q}_{\mathrm{AP}}$ follows (almost immediately) from the irreducibility of $Q_{\mathrm{AP}}$.
In order to prove this, we first argue that a strictly positive transition rate in the detailed state space description induces a strictly positive transition rate in the reduced state space description.
That is, for any $x, y \in \mathscr{X}_{\mathrm{det}}$, $Q_{\mathrm{AP}}(x, y) > 0$ implies that $\tilde{Q}_{\mathrm{AP}}(\Gamma(x), \Gamma(y)) > 0$.
The reason for this is that any strictly positive transition rate in the detailed state space corresponds to the arrival or departure of a type 1 or type 2 request, and that in the reduced state space the arrival or departure of a type 1 or type 2 request also corresponds to a transition with non-zero rate.
While this argument is rather intuitive, it can easily be verified more formally.
Take for instance the RA policy.
One can immediately verify that for any generic state $x \in \mathscr{X}_{\mathrm{det}}$ and all $y \in \mathscr{X}_{\mathrm{det}}$ such that $Q_{\mathrm{RA}}(x, y) > 0$, $\tilde{Q}_{\mathrm{RA}}(\Gamma(x), \Gamma(y)) > 0$.

Now that we have established this implication, proving the irreducibility of $\tilde{Q}_{\mathrm{AP}}$ becomes almost trivial.
We consider the RA policy, but the reasoning for the LF and MF policies is entirely similar.
Fix any two distinct states $x'$ and $y'$ in the reduced state space $\mathscr{X}_{\mathrm{red}}$.
Because of the surjectivity of the map $\Gamma$ between both state spaces, there are some $x, y \in \mathscr{X}_{\mathrm{det}}$ such that $\Gamma(x) = x'$, $\Gamma(y) = y'$ and $x \neq y$.
As $Q_{\mathrm{RA}}$ is irreducible, there exists a sequence $x = x_0, \dots, x_{n} = y$ in $\mathscr{X}_{\mathrm{RA}}$ such that for any $i \in \{ 1, \dots, n \}$, $Q_{\mathrm{RA}}(x_{i-1}, x_{i}) > 0$.
By the argument made above, this implies that $\tilde{Q}_{\mathrm{RA}}(\Gamma(x_{i-1}), \Gamma(x_{i})) > 0$ for all $i \in \{ 1, \dots, n \}$.
Recall that by construction $\Gamma(x_{0}) = x'$ and $\Gamma(x_{n}) = y'$, such that we can infer from this that $y'$ is accessible from $x'$.
As $x'$ and $y'$ were arbitrary distinct states in $\mathscr{X}_{\mathrm{red}}$, this proves that $\tilde{Q}_{\mathrm{RA}}$ is indeed irreducible.

	\section{A Closer Look at the Lower Transition Rate Operators Of Interest}
\label{app:LTROAndLumping}
From a practical point of view, an important question is how to efficiently execute the minimisation in \eqref{eq:Qlower}.
To discuss this, we fix some $x = (i, j, e) \in \mathscr{X}_{\mathrm{red}}$ and define
\begin{align*}
	x_{1,+}^{=} &\coloneqq (i+1, j, e), & x_{1,+}^{-} &\coloneqq (i+1, j, e-1), \\
	x_{1,-}^{=} &\coloneqq (i-1, j, e), & x_{1,-}^{+} &\coloneqq (i-1, j, e+1), \\
	x_{2,+} &\coloneqq (i, j+1, e-1), & x_{2,-} &\coloneqq (i, j-1, e+1).
\end{align*}
For convenience' sake, we first assume that all these states are feasible, or equivalently that all these states are elements of the reduced state space $\mathscr{X}_{\mathrm{red}}$.
If we use AP as a place holder for RA, LF or MF, it holds that
\begin{align*}
	[\underline{Q}_{\mathrm{AP}} f](x)
	&= \min_{Q \in \mathscr{Q}_{\mathrm{AP}}} [Q f](x) \\
	&= \min \big\{ -(\lambda_{1} + \lambda_{2} + i \mu_{1}  + j \mu_{2}) f(x) \\
	&\qquad\qquad + \lambda_{2} f(x_{2,+}) + j \mu_{2} f(x_{2,-}) \\
	&\qquad\qquad + (\lambda_{1} - \lambda_{\mathrm{AP}}^{-}) f(x_{1,+}^{=}) + \lambda_{\mathrm{AP}}^{-} f(x_{1,+}^{-}) \\
	&\qquad\qquad + (i - i^{+}) \mu_{1} f(x_{1,-}^{=}) + i^{+} \mu_{1} f(x_{1,-}^{+}) \\
	&\hspace{12em} \colon i^{+} \in [i_{\min}, i_{\max}] \big\},
\end{align*}
where $\lambda_{\mathrm{AP}}^{-}$, $i_{\min}$ and $i_{\max}$ are functions of $(i,j,e)$ as defined in Section~\ref{ssec:ApprMCMod:TransitionRates}.
The solution to the resulting minimisation is trivial: the minimising value of $i^{+}$ is either $i_{\min}$ or $i_{\max}$.
The edge cases where one or more of the six states above are not feasible can be treated similarly.

Computing such minima in practice can be done in multiple ways.
In our numerical computations, we a priori construct two sparse---because every row has at most 7 non-zero elements---transition rate matrices $Q_{\min}$ and $Q_{\max}$: for $Q_{\min}$ we use $i_{\min}$ for all rows and for $Q_{\max}$ we use $i_{\max}$ for all rows.
It then trivially holds that
$
	[\underline{Q}_{\mathrm{AP}} f](x)
	= \min \{ [Q_{\min} f](x), [Q_{\max} f](x)  \}.
$
If one is only interested in the $x$-component of $\underline{Q}_{\mathrm{AP}} f$, then one simply computes the dot product of the $x$-th row of $Q_{\min}$ and $Q_{\max}$ with $f$ and selects the minimum of the two obtained values.
If one is interested in all components of $\underline{Q}_{\mathrm{AP}} f$, then one computes the product of each of the two matrices $Q_{\min}$ and $Q_{\max}$ with the column vector $f$, and computes the component-wise minimum of the two obtained column vectors.

The policy-independent case is very similar to the policy-dependent case.
Under similar assumptions on $x = (i, j, e)$ as for the policy-dependent case, the minimisation in \eqref{eq:Qlower} reduces to
\begin{align*}
	[\underline{Q}_{\mathrm{PI}} f](x)
	&= \min_{Q \in \mathscr{Q}_{\mathrm{PI}}} [Q f](x) \\
	&= \min \big\{ -(\lambda_{1} + \lambda_{2} + i \mu_{1} + j \mu_{2}) f(x) \\
	&\qquad\qquad + \lambda_{2} f(x_{2,+}) + j \mu_{2} f(x_{2,-})   \\
	&\qquad\qquad + (1 - \alpha) \lambda_{1} f(x_{1,+}^{=}) + \alpha \lambda_{1} f(x_{1,+}^{-}) \\
	&\qquad\qquad + (i - i^{+}) \mu_{1} f(x_{1,-}^{=}) + i^{+} \mu_{1} f(x_{1,-}^{+}) \\
	&\hspace{7em} \colon i^{+} \in [i_{\min} , i_{\max}], \alpha \in [0, 1]  \big\}.
\end{align*}
Note that the cases $e = 0$ and $i = n_{2} (m_{2} - j - e)$ cannot occur under the assumptions we made.
Solving the above minimisation is again trivial, as we can independently choose the minimising values of $i^{+}$ and $\alpha$.
Here too, the edge cases where one or more of the six adjacent states are not feasible can be treated similarly.
In our numerical computations, we execute the required minimisations by determining the component-wise minimum of four column vectors.
These four vectors are obtained by multiplying four (sparse) transition rate matrices---one for every combination of extremal values of $i^{+}$ and $\alpha$---with the column vector $f$.

	\section{Lower Transition Rate Operators}
\label{app:LTROs}

Let $\mathscr{R}$ denote the set of all transition rate matrices, and let $\mathscr{Q} \subset \mathscr{R}$ be some set of transition rate matrices.
For all $x \in \mathscr{X}$, we let $\mathscr{Q}_{x}$ denote the set of $x$-rows in $\mathscr{Q}$:
\[
	\mathscr{Q}_{x} \coloneqq \{ Q(x, \cdot) \colon Q \in \mathscr{Q} \},
\]
where $Q(x, \cdot)$ denotes the $x$-row of $Q$.
Then \citet[Definition~7.3]{2016Krak} say that $\mathscr{Q}$ has \emph{separately specified rows} if
\[
	\mathscr{Q}
	= \{ Q \in \mathscr{R} \colon (\forall x \in \mathscr{X}) ~ Q(x, \cdot) \in \mathscr{Q}_{x} \}.
\]
In words, we require that if we construct a transition rate matrix $Q$ by selecting an arbitrary $x$-row $Q(x, \cdot)$ in the set of $x$-rows $\mathscr{Q}_{x}$ independently for each row, then $Q$ is contained in $\mathscr{Q}$.

The sets $\mathscr{Q}_{\mathrm{RA}}$, $\mathscr{Q}_{\mathrm{LM}}$ and $\mathscr{Q}_{\mathrm{PI}}$ of transition rate matrices that are compatible with the relevant (bounds on the) transition rates have separately specified rows.
Indeed, the only transitions with linked rates---in this case the departure (or arrival) of a type 1 message---start in the same state, such that the rates of a compatible transition rate matrix can be chosen on a per-row basis.
Therefore, the sets $\mathscr{Q}_{\mathrm{RA}}$, $\mathscr{Q}_{\mathrm{LM}}$ and $\mathscr{Q}_{\mathrm{PI}}$ can be used to define lower transition rate operators $\underline{Q}_{\mathrm{RA}}$, $\underline{Q}_{\mathrm{LM}}$ and $\underline{Q}_{\mathrm{PI}}$ according to \eqref{eq:Qlower}, and these are guaranteed to satisfy \eqref{eqn:LimitComputationOfTheLowerTransitionOperator}.

	\subsection{Ergodicity}
\label{app:ImpreciseErgodicity}
As explained in \cite{2016DeBock}, a lower transition rate operator $\underline{Q}$ is ergodic if $\lim_{t \to +\infty} [\underline{T}_{s}^{t} f](x_{s})$ is independent of the initial state $x_{s} \in \mathscr{X}$.
De Bock provides a necessary and sufficient condition for the ergodicity of a lower transition rate operator in \cite[Theorem~19]{2016DeBock}.
In order to state this result, we first need to introduce some notation and terminology.

Similarly to the precise case, we define a binary accessibility relation $\cdot \upacces \cdot$.
Let $\underline{Q}$ be a lower transition rate operator with state space $\mathscr{X}$.
A state $x \in \mathscr{X}$ is said to be upper reachable from the state $y \in \mathscr{X}$, denoted by $y \upacces x$, if either (i) $x = y$, or (ii) $x \neq y$ and there is a sequence $y = x_{0}, \dots, x_{n} = x \in \mathscr{X}$ such that $[\overline{Q} \mathbb{I}_{x_{k}}](x_{k-1}) > 0$ for all $k \in \{ 1, \dots, n \}$.
In the second condition, $\overline{Q}$ denotes the upper transition rate operator, defined for all real-valued functions~$f$ on $\mathscr{X}$ by the conjugacy relation $\overline{Q} f \coloneqq -\underline{Q} (-f)$.
Note that the upper accessibility relation $\upacces$ is just the generalisation to upper transition rate operators of the accessibility relation $\acces$ for transition rate matrices.

A transition rate operator $\underline{Q}$ is called \emph{top class regular} if the top class $\mathscr{X}_{\mathrm{top}}$ is non-empty, where
\newcommand{\upperreach}{\xrightarrow{\cdot}}
\newcommand{\lowerreach}{\xrightarrow[\cdot]{}}
\[
	\mathscr{X}_{\mathrm{top}}
	\coloneqq \{ x \in \mathscr{X} \colon (\forall y \in \mathscr{X})
	~y \upacces x \}.
\]
In the imprecise case, however, top class regularity is not a sufficient condition for the ergodicity of a lower transition rate operator \cite{2016DeBock}.
A second condition called \emph{top class absorption} is also necessary.
However, if the top class $\mathscr{X}_{\mathrm{top}}$ is equal to the state space $\mathscr{X}$, then top class absorption is trivially satisfied.
Hence, irreducibility---$\mathscr{X}_{\mathrm{top}} = \mathscr{X}$---is a sufficient condition for ergodicity.

	\subsection{Irreducibility of the Lower Transition Rate Operators of Interest}
The irreducibility of the policy-dependent lower transition rate operators $\underline{Q}_{\mathrm{RA}}$ and $\underline{Q}_{\mathrm{LM}}$, as well as the irreducibility of the policy-independent lower transition rate operator $\underline{Q}_{\mathrm{PI}}$, follows from the irreducibility of the transition rate matrices $\tilde{Q}_{\mathrm{RA}}$ and $\tilde{Q}_{\mathrm{LM}}$.

Fix any two distinct states $x$ and $y$ in $\mathscr{X}_{\mathrm{red}}$.
By the irreducibility of $\tilde{Q}_{\mathrm{RA}}$, there exists a sequence $y = x_{0}, \dots, x_{n} = x$ in $\mathscr{X}_{\mathrm{red}}$ such that $\tilde{Q}_{\mathrm{RA}}(x_{i-1}, x_{i}) > 0$ for all $i \in \{ 1, \dots, n \}$.
Note that as $\tilde{Q}_{\mathrm{RA}} \in \mathscr{Q}_{\mathrm{RA}}$, it follows from \eqref{eq:Qlower} and the conjugacy relation between $\underline{Q}_{\mathrm{RA}}$ and $\overline{Q}_{\mathrm{RA}}$ that, for any $i \in \{ 1, \dots, n \}$,
\begin{multline*}
	[\overline{Q}_{\mathrm{RA}} \mathbb{I}_{x_{i}}](x_{i-1})
	= - [\underline{Q}_{\mathrm{RA}} (- \mathbb{I}_{x_{i}})](x_{i-1}) \\
	\geq - [\tilde{Q}_{\mathrm{RA}} (- \mathbb{I}_{x_{i}})](x_{i-1})
	= \tilde{Q}_{\mathrm{RA}}(x_{i-1}, x_{i}).
\end{multline*}
This inequality implies that $[\overline{Q}_{\mathrm{RA}} \mathbb{I}_{x_{i}}](x_{i-1}) > 0$ for all $i \in \{ 1, \dots, n \}$, or equivalently that $x$ is upper accessible from $y$.
As $x$ and $y$ were arbitrary distinct states, we conclude that $\underline{Q}_{\mathrm{RA}}$ is irreducible.
Entirely similar reasoning can be used to prove the irreducibility of $\underline{Q}_{\mathrm{LM}}$ and that of $\underline{Q}_{\mathrm{PI}}$

\end{document}